\documentclass[useAMS,usenatbib]{mn2e}
\usepackage{graphicx}
\usepackage{subfigure}
\usepackage{amsmath}
\usepackage{amssymb}
\usepackage{color}
\usepackage{xcolor}
\usepackage{natbib}
\usepackage{slantsc}

\title[]{The variation in molecular gas depletion time among nearby galaxies: II
the impact of galaxy internal structures}
\author[M.-L. Huang et al.]{Mei-Ling Huang$^{1}$\thanks{E-mail:
mlhuang@mpa-garching.mpg.de}, Guinevere Kauffmann$^{1}$\\
$^{1}$Max-Planck Institute for Astrophysics, Karl-Schwarzschild-Str. 1, D-85748 Garching, Germany\\
}
\begin{document}
\date{ in original form 2014 Sep}
\pagerange{\pageref{firstpage}--\pageref{lastpage}} \pubyear{2002}
\maketitle
\label{firstpage}

\begin{abstract}
We use a data set of nearby galaxies drawn from the HERACLES, 
ATLAS$^{3D}$, and COLD GASS surveys to study variations in molecular gas 
depletion time (t$_{dep}$) in galaxy structures such as bulges, grand-design 
spiral arms, bars and rings. Molecular gas is traced by CO 
line emission and star formation rate (SFR) is derived using the combination of 
far-ultraviolet and mid-infrared (MIR) data. The contribution of old stars to 
MIR emission for the ATLAS$^{3D}$ sample is corrected using 2MASS K-band images. 
We apply a two-dimensional image decomposition algorithm to decompose galaxies
into bulges and discs. Spiral arms, bars and rings are identified in the
residual maps, and molecular gas depletion times are derived on a square
grid of 1 kpc$^2$ size. In previous work, we showed that  
t$_{dep}$ correlates strongly with specific star formation
rate (sSFR).  We now find that at a given sSFR, the bulge has shorter
 t$_{dep}$ than the disc. The shift to shorter depletion times is most
pronounced in the inner bulge ($R < 0.1R_e$). Grids from galaxies with bars
and rings are similar to those from galactic bulges in that they have
reduced t$_{dep}$ at a given sSFR.  In contrast, the t$_{dep}$ versus sSFR
relation in the discs of galaxies with spiral arms is displaced to longer
t$_{dep}$ at fixed sSFR. We then show that the differences  in the
t$_{dep}-$sSFR relation for bulges, discs, arms, bars and rings can be
linked to variations in {\em stellar}, rather than gas surface density
between different structures. Our best current predictor for t$_{dep}$,
both globally and for 1 kpc grids, is given by 
$t_{dep} = -0.36\log(\Sigma_{\rm SFR})-0.50\log(\Sigma_{*})+5.87$.

\end{abstract}

\section{Introduction} 												
Numerous studies have attempted to understand how gas is converted into
stars. Observational data indicate 
a power-law relation between these two quantities:
\[
\rm\Sigma_{SFR} \propto \rm\Sigma_{gas}^{N},
\]
where $\rm \Sigma_{SFR}$ and $\rm \Sigma_{gas}$ are 
the star formation rate (SFR) and gas surface densities in units of
M$_{\sun}$ yr$^{-1}$ kpc$^{-2}$ and M$_{\sun}$ pc$^{-2}$.
Such relations are often called the Kennicutt-Schmidt (K-S) laws 
(Schmidt 1959; Kennicutt 1998).

Only recently have observations provided high-resolution, multiwavelength 
images of nearby galaxies for spatially-resolved, sensitive measurements 
of $\rm \Sigma_{gas}$ and $\rm \Sigma_{SFR}$, allowing exploration of the K-S 
relation on sub-kpc scales. For instance, Bigiel et al. (2008, 2011) used 
the CO(J=2-1) line as an H$_2$ gas tracer and far-ultraviolet (FUV) plus 24$\micron$ 
emission as a SFR tracer to study the K-S 
relation at sub-kpc resolution in nearby disc galaxies observed as part of the 
HERA CO-Line Extragalactic Survey (HERACLES; Leroy et al. 2008). They proposed  a 
linear K-S relation (and thus a constant molecular gas depletion time (t$_{dep}$)), 
independent of local conditions such as orbital timescale, mid-plane gas pressure 
and disc stability parameter $Q$ (Leroy et al. 2008). Some subsequent works, however, 
reported non-linear relations between $\rm \Sigma_{SFR}$ and $\rm \Sigma_{H_{2}}$   
using different fitting methods (Shetty et al. 2013) or
new data (e.g., Momose et al. 2013; Pan et al. 2014). 

\citet{sanb} investigated the relation between integrated 
molecular gas depletion time and global galaxy parameters statistically using galaxies  
observed as part of the COLD 
GASS project, which obtained IRAM $\sim$22$\arcsec$ single beam CO(1-0) line
measurements for  $\sim$360 galaxies with stellar masses 
$\rm 10^{10}-10^{11.5} M_{\sun}$ in the redshift range 0.02 -- 0.05 \citep{sana}.  
SFR was estimated  by fitting stellar population 
models to optical and ultraviolet (UV) broad-band photometry. An estimate of the  
molecular gas depletion time of the full galaxy was made  by applying aperture 
corrections to the CO line luminosities.  Molecular gas depletion time 
was found to correlate well with various galaxy parameters such as stellar mass, stellar 
surface density, concentration of the light (i.e., bulge-to-disc ratio), 
near-UV (NUV)$-$r color and specific star formation rate (sSFR; defined as 
the ratio of SFR to stellar mass).

We subsequently improved on this work by estimating SFR with the combination 
of GALEX FUV and WISE 22$\micron$ data within fixed apertures matched to the
22$\arcsec$ beam size of the gas observations \citep{hua}.
Dependences of the depletion time on galaxy structural parameters such as
stellar surface density and concentration index were then found to be 
much weaker, or even absent.
We went on to demonstrate that the {\em primary dependence} of t$_{dep}$ is on sSFR.
All other remaining correlations, such as the t$_{dep}$--stellar mass relation, 
were demonstrated to be secondary, i.e., induced by the fact that sSFR in turn 
correlates with a variety of other global galaxy parameters. We note that 
a comprehensive study based on 500 star forming galaxies at redshifts 
from z = 0 to 3 also supports our findings (Genzel et al. 2014). 

We also compared the results obtained from the COLD GASS galaxies 
to estimates of $t_{dep}$ on 1 kpc scales using high-resolution
CO and SFR maps from the  HERACLES survey. The global t$_{dep}$--sSFR 
relation derived from the COLD GASS sample and that derived 
from 1 kpc grids placed on HERACLES galaxies agree remarkably well.
This suggests that the local molecular gas depletion time is dependent on the local
fraction of young-to-old stars and that galaxies with high current-to-past-averaged
star formation activity, will consume their molecular gas reservoir sooner.

Another aspect that has not been fully understood is the role of galaxy 
structures such as bulges, arms, bars, or rings in determining 
the molecular gas depletion time.  
We note that simply using all available grids from the whole galaxy or integrated 
values of gas mass and SFR to study molecular gas depletion times   
might smooth out variations caused by different structures.

Many studies have investigated whether the 
star formation efficiency (SFE) varies in differnt galactic environments.
For instance, Meidt et al. (2013) studied the inner region of the spiral galaxy
M51 with a high resolution CO (1-0) map ($\sim$40 pc) and proposed that streaming 
motions, induced by gravitational instabilities
due to bars and arms, cause variations in gas depletion time. 
Momose et al. (2010) observed a barred spiral galaxy, NGC 4303, with CO (1-0) 
observations and found that its arms have twice higher SFEs than its bar, although 
strong CO emission is both seen in the bar and the disc region along the arms.
Fujimoto et al. (2014) proposed that these results could be understood in a model
where star formation is regulated by collisions between molecular clouds.
However, other studies have come to the opposite conclusion
about the viability of the cloud collision model 
(Foyle et al. 2010; Eden et al. 2012).
Other observational studies using CO(1-0) line luminosities as the gas tracer
and H$\alpha$ or 24$\micron$ luminosities as the SFR tracer 
have demonstrated that the SFE in spiral arms is enhanced compared to 
regions outside the arms (e.g., Lord \& Young 1990;
Knapen et al. 1996; Rebolledo et al. 2012; Hirota et al. 2014).

We note that most of these studies explored the influence of structures on 
SFE based on samples of one or only a few galaxies, and they did not consider 
variations in SFE with respect to the t$_{dep}-$sSFR relation. 
In this paper, we try to understand whether galaxy structures induce significant
variation in molecular gas depletion time by systematically separating the grids into 
those in bulge, arm, bar, and ring regions, and by studying how the local depletion 
time from these  different regions varies with respect to the 
established t$_{dep}-$sSFR relation.

Our paper is organized as follows. We describe the multiwavelength data sets 
used in this work in Section 2, and our methods for identifying galaxy 
structures and deriving t$_{dep}$ and sSFR on local scales in Section 3. 
Our results are presented in Section 4, 
where we look at the t$_{dep}$--sSFR relations for the grids in the bulge, 
arm, bar, and ring regions respectively.  
Finally, we compare our results with previous studies and discuss our findings 
in Section 5.

\section{Data}    													
In this section we introduce the data sets we use in this work. 
The HERACLES sample provides spatially-resolved CO maps with
resolution  $\lesssim$ 1kpc, appropriate for pixel-by-pixel studies.
Because of this good resolution, we are able to identify the pixels lying in  
galaxy structures such as bulges or spiral arms. ATLAS$^{3D}$ galaxies are mostly 
early-type galaxies and are included because this data supplements
the number of data points we are able to extract from the bulge-dominated
regions of galaxies. The COLD GASS project targets more distant sources and 
therefore provides only integrated values of the molecular gas mass. 
Nevertheless, the  large sample size gives us the opportunity to carry 
out a statistical analysis of global relations between 
gas depletion time and other galaxy properties.

\subsection{HERACLES}  												
To study molecular gas depletion time on sub-kpc scales, we use public 
data from the HERA CO-Line Extragalactic Survey (HERACLES; \citealt{ler08}).
HERACLES has released CO ($J=2-1$) maps for 48 nearby galaxies,
achieving a spatial resolution $\sim$13 arcsec and an 
average H$_{2}$ surface density detection limit of $\sim$3 M$_{\sun}$pc$^{-2}$.
We analyze  21 massive galaxies with log (M$_*$/M$_{\sun}$) $\geq$ 10 from the catalog 
in \citet{ler13} so that we avoid those low-mass galaxies with less reliable  
$H_2$ mass measurements. 
All our selected galaxies are located within a distance of $\sim$20 Mpc.

A variety of ancillary data is available for the HERACLES sample,
which allow us to derive the SFR and stellar mass surface densities.   
The relevant data includes FUV images from GALEX All-sky Imaging survey
(AIS) and Nearby Galaxy Survey (NGS; \citealt{gil}), 24 $\micron$ data from the 
Spitzer Infrared Nearby Galaxies Survey (SINGS; \citealt{ken03}),
H\textsc{i} maps from The H\textsc{i} Nearby Galaxies Survey 
(THINGS; \citealt{wal}), and optical images from Sloan Digital Sky Survey (SDSS). 
FUV images from GALEX satellite have effective wavelength 1528 \AA\  and 
angular resolution $\sim$4.3 arcsec FWHM. 
SINGS provides MIPS 24$\mu$m images with an angular resolution $\sim$6 arcsec
and 3 $\sigma$ sensitivity $\sim$0.21 MJy sr$^{-1}$. 
The maps (natural-weighting) from THINGS have an angular resolution 
of $\sim$11 arcsec and are sensitive to $\rm \Sigma_{HI}$ 
$\geq$ 0.5 M$_{\sun}$ pc$^{-2}$.

Combining these public data sets, we estimate gas, SFR, and stellar mass surface 
densities as a function of position in the galaxy by dividing the galaxy into a 
set of square cells with 1 kpc$^2$ size. In this way, 
we are able to separate the grids within bulges, arms, 
bars, rings from those in the rest of the galaxy. Galaxy parameters such as 
distance, inclination angle, position angle are taken directly from \citet{ler13}.

\subsection{ATLAS$^{3D}$}  												
To increase the amount of data in bulge regions, we make use of a subset of
the galaxies observed as part of the  ATLAS$^{3D}$ project \citep{cap}, 
with publicly-available resolved interferometric CO(1-0) maps \citep{ala}.
The  full ATLAS$^{3D}$ sample consists of a volume-limited sample of 260 nearby early-type 
(elliptical E and lenticular S) galaxies (ETGs) with a cut-off absolute magnitude 
of M$_K$ = $-$21.5. The ATLAS$^{3D}$ galaxies  were observed in CO(1-0) and CO(2-1) 
by the IRAM 30m telescope \citep{you}, and those with detected CO integrated fluxes 
larger than 19 Jy kms$^{-1}$ were  imaged using the
Combined Array for Research in Millimeter Astronomy (CARMA) \citep{ala}. 
This data set includes 30 ETGs. Together with 10 CO-detected galaxies that
have interferometric CO(1-0) data from the literature, there are a total 
40 ETGs with resolved CO maps in the ATLAS$^{3D}$ sample.
The reader is referred to \citet{ala} for further details about the sample, 
the observations and the analyses. 

We derive SFR using the combination of FUV emission from GALEX data and mid-infrared 
emission at 22\micron\ from WISE data. The FUV maps are retrieved from GALEX Data 
Release 7 products.  When several atlas images from different observing programs 
are available, we always use the one with the longest exposure time. 
The exposure time for $\sim$2/3 of the sample is longer than 1500 seconds; 
the rest of galaxies have data from the AIS program with typical exposure 
time $\sim$100 -- 200 seconds and limiting magnitude $\sim$20.5 AB magnitude. 
There are 3 galaxies without FUV observations from GALEX. 
WISE provides  22\micron\  images of the whole sky, with angular resolution 
12 arcsec and  5$\sigma$ point-source sensitivity $\sim$ 6mJy.
All galaxies have WISE observations.
The stellar mass is derived from SDSS 5-band images using a SED-fitting method;
the SDSS images are downloaded from SDSS Data Release 9. 

In Figure 1 we compare the  basic properties of the galaxies in the
 HERACLES and ATLAS$^{3D}$ samples. Generally, their stellar masses 
 are comparable. The HERACLES galaxies are much closer than the 
 ATLAS$^{3D}$ galaxies, so the angular 
disc sizes (R$_{25}$) of the former sample are larger than the latter sample. 
The distributions of the  concentration index, R$_{90}$/R$_{50}$, which 
is tightly related to the bulge-to-disc
ratio of the galaxy (\citealt{gad}; \citealt{wei}), reflect the fact that  
the ATLAS$^{3D}$ galaxies are mostly ETGs and the HERACLES sample
consists of mostly disc galaxies. 

\begin{figure} 														
\begin{center}
 \includegraphics[scale=0.42]{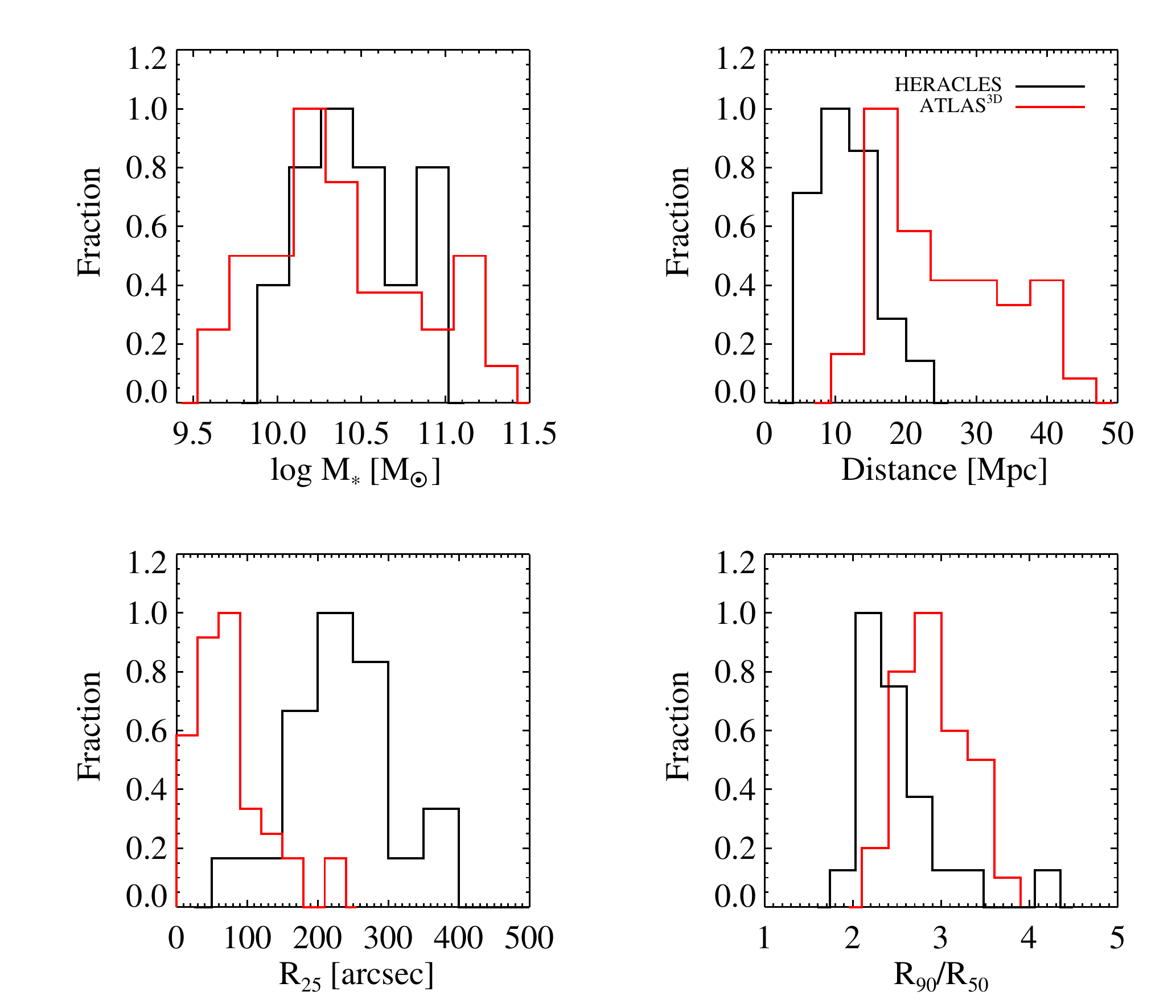}
    \caption{Comparison of stellar mass, distance, B-band isophotal radius at 25 mag 
	  arcsec$^{-2}$ (R$_{25}$), concentration index (R$_{90}$/R$_{50}$) for HERACLES 
	  galaxies (black line) and ATLAS$^{3D}$ galaxies (red line). }
  \label{f}
\end{center}
\end{figure}

\subsection{COLD GASS} 												
The sample is drawn from the COLD GASS survey catalogue \citep{sana, sanb,sanc}, 
which contains CO ($J=1-0$) line measurements from the IRAM 30m telescope for 
$\sim$360 nearby galaxies with stellar masses in the range  $10^{10}-10^{11.5} M_{\sun}$ 
and redshifts in the range $0.025 < z < 0.05$. We only select the galaxies with 
detected CO fluxes in our studies. The reader is referred to \citet{sana} for 
a detailed description of the sample selection and the observations. 

In the same fashion as was done for the  ATLAS$^{3D}$ sample, we derive SFR by summing  
GALEX FUV and WISE 22$\micron$\ luminosities and we estimate stellar mass
by fitting stellar population synthesis models to SDSS 5-band images.
We retrieve the FUV maps from GALEX Data Release 7 products;  
the maps with the longest exposure time are always preferred.
Most of our FUV images are from the Medium Imaging survey (MIS) with 
typical exposure time $\sim$1500 seconds;  81 out of
366 galaxies only have data from the AIS program with typical exposure 
time $\sim$100 -- 200 seconds. The limiting magnitude for 
MIS is 23.5 AB magnitude and for AIS, it is  20.5 AB magnitude.
All galaxies have both WISE and SDSS imaging data.

\begin{figure} 
\begin{center}
 \includegraphics[scale=0.35]{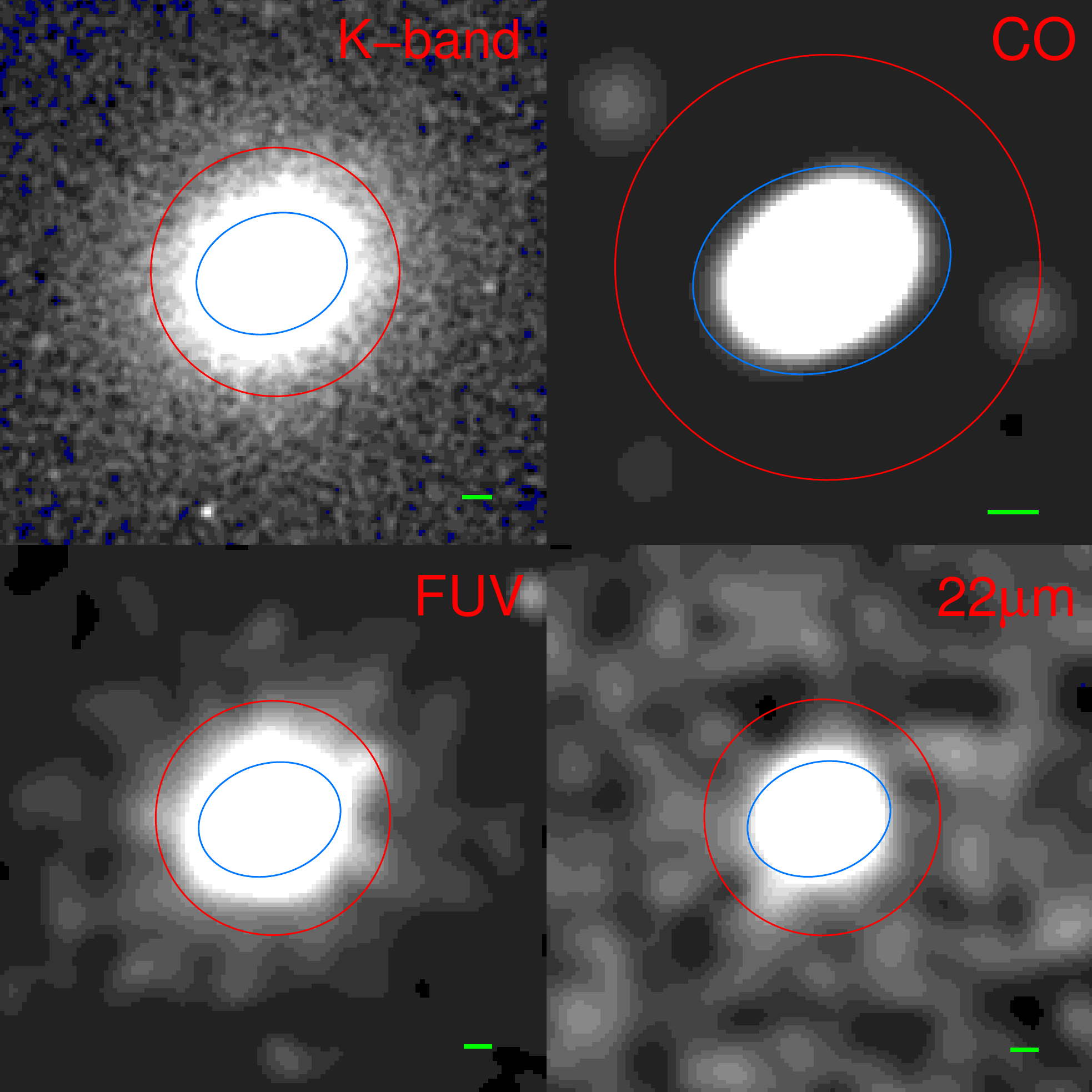}
   \caption{2MASS K-band (upper left), CARMA CO (upper right), GALEX FUV (lower left) 
	        and WISE 22\micron\ (lower right) maps for NGC3607 in the ATLAS$^{3D}$ sample. 
			All maps are convolved to 13 arcsec resolution.
			The blue ellipses and red circles indicate the aperture derived from CO(1-0) map and
			the half-light radius R$_{e}$ in K-band taken from Cappellari et al. (2011).  
		    The green bar shows the physical scale of 1kpc in each map.}
  \label{f2}
\end{center}
\end{figure}

\begin{figure} 													          
\begin{center}
 \includegraphics[scale=0.4]{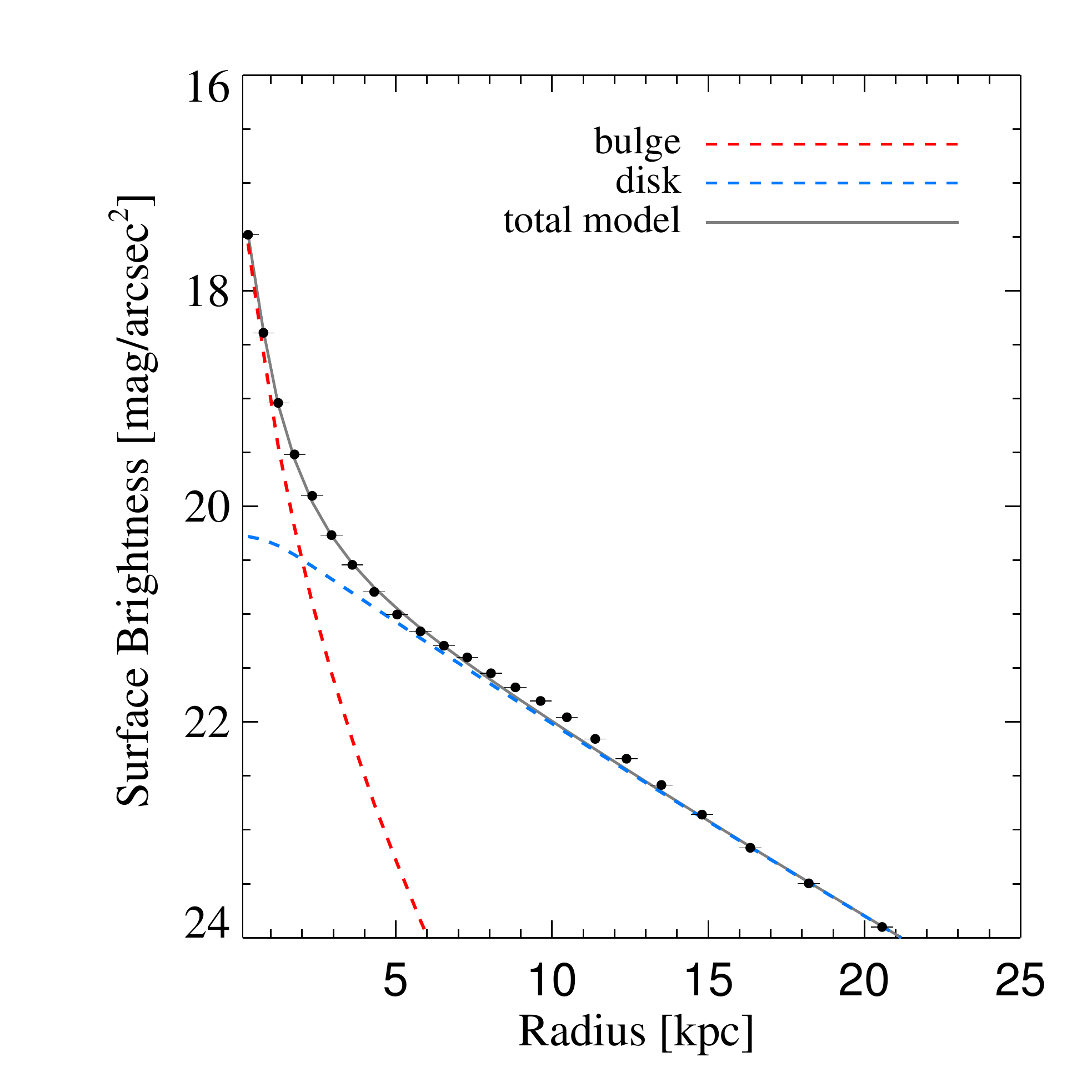}
    \caption{An example of a  one-dimensional decomposition profile for a galaxy in the HERACLES sample. 
	  Black points show  the mean surface brightness of the g-band image in elliptical annuli
      with the axial ratio and position angle drawn from the disc model.
	  The grey line denotes the total model flux. The red and blue dashed lines denote the model fluxes 
	  for the bulge and the disc.  }
  \label{f3}
\end{center}
\end{figure}

\section{Methods}  												
Molecular gas depletion time is defined as $\Sigma_{\rm H_2}$/$\Sigma_{\rm SFR}$ 
or M$_{\rm H_2}$/SFR, where $\Sigma_{\rm H2}$ and $\Sigma_{\rm SFR}$ are molecular 
gas and SFR surface densities, and M$_{\rm H2}$ is molecular gas mass. 
Specific SFR (sSFR) is defined as SFR/M$_{*}$ or 
$\Sigma_{\rm SFR}$/$\Sigma_{*}$, where $\Sigma_{*}$ is stellar mass surface density.
We use the first definition for the spatially-resolved HERACLES sample; 
the second definition is used for the ATLAS$^{3D}$ and COLD GASS samples to derive
an integrated depletion time and sSFR.

\subsection{Overview of the spatial scales probed by the observations}                                          
For the HERACLES sample, the spatial resolution of the CO maps is $\sim$13\arcsec\ , which is 
$\lesssim$1kpc. The GALEX FUV and Spitzer 24$\micron$ maps 
have better resolution, so we degrade them to the same resolution 
as the CO maps. The sizes of structures we want to study,
such as bulges, arms, bars and rings, are  well-resolved with 1kpc smoothing.
We therefore use 1kpc $\times$ 1kpc grid cells as our sampling elements.

ATLAS$^{3D}$ includes a subset of $\sim$40 ETGs with resolved CO maps and supplements 
data for bulge regions. Given that the size of the CO-emitting regions is only a few 
kpcs or even smaller and that the gas surface density sensitivity limit varies from one galaxy 
to another for the ATLAS$^{3D}$ project, we do not derive the depletion time within 
1kpc$^{2}$ grids, but within the CO-emitting region. We first smooth CO maps to the 
same 13\arcsec\ resolution as the  WISE image. We run SEXTRACTOR 
on the CO moment0 maps and use the Kron aperture as the size of CO-emitting region.
The CO flux within the Kron aperture is converted to the molecular gas mass
by applying a Galactic CO-to-H$_{2}$ conversion factor.
We note that we measure the SFR and the stellar mass using the same aperture 
as the CO emission. 
An example of our multi-band images with the aperture derived 
from the CO map is shown in Figure 2.

The CO emission for the  COLD GASS galaxies is observed within the 
IRAM 22$\arcsec$ Gaussian beam.
Once again, SFR must be measured within the same region where molecular gas is observed; thus, 
we derive SFR for the COLD GASS galaxies by placing a 22\arcsec\ Gaussian beam
on the FUV and 22$\micron$ maps. As pointed out in \citet{hua}, the depletion 
timescales that we derive for the COLD GASS galaxies 
agree better with the HERACLES results than those in Saintonge et al (2011b). 

\begin{figure*} 													     
\begin{center}
 \includegraphics[scale=0.78]{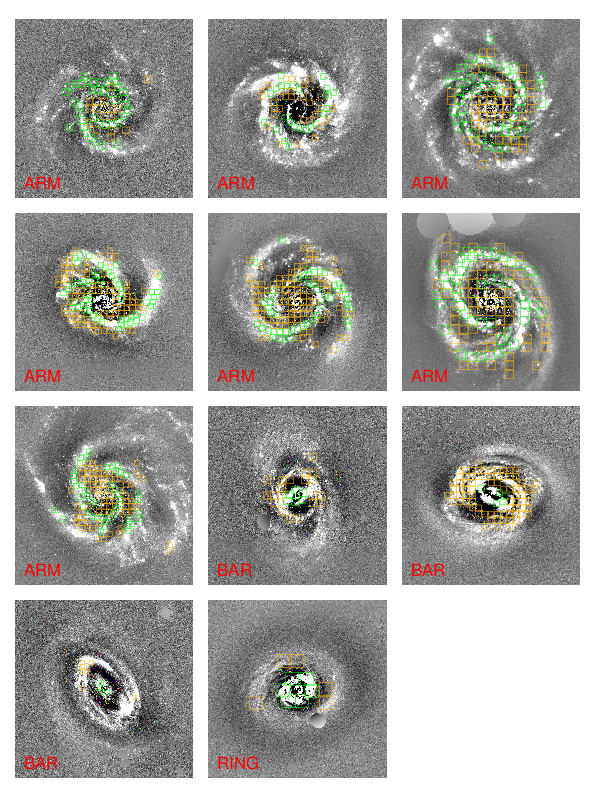}
    \caption{SDSS g-band residual maps for HERACLES galaxies with grand-design spiral 
	  arms, bars and rings. Coloured squares show grids of 1-kpc$^{2}$ size in the area of 
		the targeted structure (green) and in the rest of the galaxy
              (yellow) . Note that we only show the 
	    grids that lie above the gas and SFR sensitivity limits.}
  \label{f4}
\end{center}
\end{figure*}

\subsection{Identification of bulges/arms/bars/rings} 					     
In this section, we fit S\'{e}rsic models to SDSS g-band images and identify 
bars/rings/spiral arms from the residual images.  After the main features of 
galaxies, i.e. the bulge and the disc, are removed, the residual maps can reveal fine 
structures such as rings or arms more clearly.

We first mask stars and background galaxies in the  SDSS images 
using the colour-based masks described in \citep{mun}. We inspect the images again
and mask stars by hand when necessary. 
The masked regions are marked as bad pixels which will not be used for fitting.
The maps of \citet{sch} are used to correct for Galactic extinction.
We next apply a two-dimensional image decomposition program called GALFIT \citep{pen} 
to the SDSS g-band images to decompose galaxies into bulges and disks. To fit the 
luminosity profile of the galaxies, the code assumes a two-component model, 
S\'{e}rsic bulge plus S\'{e}rsic disk profile, where the S\'{e}rsic $n$ values for 
the bulge and disk are allowed to vary from 1.5 to 4 and from 0.8 to 1.2, respectively. 
The grids within the effective radius of the bulge model are designated as the 
bulge grids. One example of the fitting result is shown in Figure 3. 
The surface brightness profile of the galaxy is clearly separated  into two components.
The effective radii of the bulge and disc are 0.8 and 8.0 kpc.

We identify seven galaxies with large arms, three barred galaxies, 
and one galaxy with a ring from the residual maps. We visually select grids within
the arm, bar and ring regions. 
Examples  are shown in Figure 4.
Note that we exclude those grids with $\rm \Sigma_{H_{2}}$ below the CO detection sensitivity limit
or  $\rm \Sigma_{SFR} < 10^{-3}$ M$_{\sun}$ yr$^{-1}$ kpc$^{-2}$, because the  SFR calibration 
does not work for these low values (see Section 3.4.1). 
As can be seen, we are able to sample between  $50-100$ grid cells  
for each galaxy, with the exception of some of the barred galaxies and the ring galaxy.
The reason why there are only a few grids on these galaxies  
is that their $\rm \Sigma_{H_{2}}$ values are generally low and thus many grids are discarded.

\subsection{Derivation of $M_{\rm H_2}$ or $\Sigma_{\rm H_2}$} 				      
For the resolved HERACLES maps,
molecular gas depletion time is defined as $\rm \Sigma_{\rm H_{2}}/ \Sigma_{\rm SFR}$.
We calculate the molecular gas depletion time in 1-kpc $\times$ 1-kpc grid cells. 
The fluxes from the reduced HERACLES CO maps are converted to $\rm \Sigma_{\rm H_{2}}$ 
in each 1-kpc grid cell adopting the Galactic CO-to-H$_2$ conversion factor, 
4.35 M$_{\sun}$ pc$^{-2}$ (K km s$^{-1}$)$^{-1}$.
Note that we discard the bins with $\rm \Sigma_{\rm H_{2}}$ $<$  3 M$_{\sun}$ pc$^{-2}$,
which is the sensitivity limit for the HERACLES CO maps.

For the COLD GASS galaxies, the observed CO 
fluxes in the catalog of \citet{sanb} are converted to the molecular gas mass. 
For ATLAS$^{3D}$, the CO flux within the Kron aperture is converted to 
the molecular gas mass.  In both cases, the Galactic CO-to-H$_2$ conversion factor 
is used.

\subsection{Derivation of SFRs or $\Sigma_{\rm SFR}$} 				            
We estimate SFR by combining the FUV and IR data.
FUV  traces the  emission from unobscured massive stars formed over the past 
$\sim$100Myr. 
UV photons absorbed by the surrounding dust
cause emission at IR wavelengths.
The linear combination of the FUV and IR emission can recover 
both the unobscured stellar emission and the dust-reprocessed emission.

\subsubsection{HERACLES}  										            
We use the FUV and 24\micron\ images to estimate  $\rm \Sigma_{SFR}$. The same
method applied to the optical images  is used to mask  stars
and background galaxies.
We re-fill the masked region with the local background. We correct the FUV images
for Galactic extinction using the maps of \citet{sch}. All the images are convolved
to a resolution of Gaussian 13\arcsec\  using the kernels released 
in \citet{ani}.  

Next we apply the method suggested in section 8.2 of \citet{ler12} to remove the 
cirrus contribution originating from evolved stellar populations. A first-order correction 
is done using the total gas surface density, i.e. the sum of $\rm \Sigma_{H_{2}}$ and 
$\rm \Sigma_{HI}$, where $\rm \Sigma_{HI}$ is obtained from THINGS data. After we 
remove this cirrus emission from the 24 $\micron$\ emission, we then adopt the updated 
calibration coefficient in \citet{ler12} to calculate the SFRs
from the linear combination of the FUV and 24 $\micron$ luminosities.
Those bins with $\Sigma\rm _{SFR}$ $<$ 10$^{-3}$ M$_{\sun}$ yr$^{-1}$ kpc$^{-2}$ are
excluded in the following analysis, because   the SFR calibration becomes poor in the
very low SFR regime \citep{ler12}.

\subsubsection{ATLAS$^{3D}$}  										           

\begin{figure} 													
\begin{center}
\includegraphics[scale=0.55]{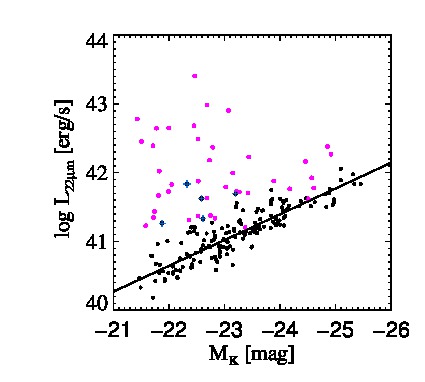}
  \caption{Relation between K-band absolute magnitude and WISE 22\micron\ luminosity.
Black and magenta solid points are the galaxies without and with CO detections from the ATLAS$^{3D}$
sample. Only  galaxies with  22\micron\ flux S/N$>$ 3 are shown. 
The black solid line is the best-fit linear regression to the CO non-detected galaxies. 
The blue plus symbols represent galaxies
excluded during the fitting process (see text).}
\end{center}
\end{figure}

We use the FUV and 22\micron\ images to estimate SFRs within  the aperture derived from the CO maps.
Both the GALEX FUV and WISE 22\micron\ images are convolved to  13\arcsec\ resolution  
using the kernels from \citet{ani}. The steps to process FUV and WISE 22\micron\ images
including star/background galaxy 
masking, sky removal, Galactic foreground extinction correction and WISE color correction,
are the same as described in \citet{hua}.  

As the ATLAS$^{3D}$  galaxies are ETGs, the contribution from old stellar 
populations to the IR emission is  significant (e.g.,Kennicutt 1998).  
\citet{dav} estimate that the average fractional contribution from evolved stars  
to the 22\micron\ luminosities of ETGs  is $\sim$25\%.
These authors found a tight correlation between K-band and 22\micron\ luminosities for
a subset of galaxies without CO detections  and
used this relation to correct the total 22\micron\ emission 
for  the contribution from old stars.

We retrieve 2MASS K-band Atlas images from the NASA/IPAC Infrared Science Archive for both 
CO-rich and CO non-detected galaxies from the  ATLAS$^{3D}$ survey. 
To remove stars or background 
galaxies, we run SEXTRACTOR and mask  detected sources, with the exception of the 
central galaxies.  We estimate the sky background as the median value of the 
pixels with values $<$ 3$\sigma$ above the median value of the whole image.
The K-band images are then convolved to the same 13\arcsec\  resolution as the
WISE 22\micron\ images. We derive  K-band and 22\micron\ luminosities for the whole galaxy 
using the Kron aperture in SEXTRACTOR. 

The  K-band absolute magnitude is plotted against 22\micron\ luminosity  for CO-rich 
and CO non-detected ETGs in Figure 5. 
As can be seen, the K-band absolute magnitude correlates well with the 22\micron\ 
luminosity for CO non-detected galaxies (black points), while there is no clear relation
for the CO-rich galaxies (magenta points).
We perform a linear fit to the CO non-detected galaxies and plot the result 
with a black solid line in Figure 5. 
The best-fit is given by   
\begin{equation}
\rmn{log} ({L_{22\mu m,\rmn{old\ stars}}\over \rmn{erg s^{-1}}})= 
 (-0.38\pm 0.01)\times M_{k}+ (32.38\pm 0.34).
\end{equation}
Note that there are 5 outliers (blue plus signs in Figure 5) with scatter larger than 0.5 dex.
They include one galaxy with apparent spiral arms, one with a star-forming ring, 
one with a high HI mass fraction, and two with either a star or small galaxy nearby.
We exclude these galaxies from our fits. 

We measure the K-band magnitudes of CO-rich ETGs within the aperture  
derived from CO maps, estimate the 22\micron\ luminosity from  old stars using 
Equation (1), and  subtract this from the total 22\micron\ emission.
We then scale the WISE 22\micron\ fluxes so that they have the same normalization 
as the MIPS 24\micron\ fluxes \citep{hua}. We convert 
the FUV and 22\micron\ luminosities  to SFRs using the same calibration given in \citet{ler12}.
\footnote {We note that the local calibrations may not be applicable to global conditions \citep{cal},
so we also try the calibration formula given in \citet{hao} which is derived for the whole galaxy.  
We find that this makes little difference to our results.}

\subsubsection{COLD GASS} 						
As in the previous section, we use GALEX FUV and WISE 22$\micron$ images to estimate SFR 
for COLD GASS galaxies within the central 22\arcsec\ region.  
The procedure is the same as mentioned above.
For consistency, we use the same calibration as we apply to HERACLES and ATLAS$^{3D}$ 
sample to convert FUV plus 22$\micron$ into SFR.  

\subsection{Derivation of stellar mass}                                         
Stellar mass is derived using the SED-fitting method in \citet{wan} by which
the stellar population synthesis models of \citet{bru} are fit to SDSS 5-band magnitudes.
The parameter, sSFR, is the SFR divided by the stellar mass.

\section{Results} 														

\subsection{The effect of internal structures on molecular gas depletion time }
\subsubsection{Bulges and discs}											
We plot the depletion time and the ratio of  SFR$_{22\mu m}$ to SFR$_{\rm FUV}$
(IR/UV) as a function of radius for bulge and disc grids in Figure 6. 
Whereas the depletion time of the bulge grids drops towards the center of the galaxy, 
the depletion time of the disc grids remains almost constant as a function of radius. 
In contrast, the IR/UV ratios of the disc and bulge grids decline from 
the center towards the outskirts -- this is not surprising because   
the IR/UV ratio is a robust indicator of dust attenuation, and  the outer 
parts of galaxies are expected to have lower metallicities and dust content. We note that the 
fact that different radial trends are found for depletion time and for IR/UV ratio
make it seem unlikely that variations in CO-to-H$_2$ conversion factor  
are responsible for the drop in the depletion time in the central bulge.              
We note that this result is consistent with results presented  in \citet{ler13} 
(see  Section 5 for further discussion). 

\begin{figure*}                                           		
\begin{center}
\includegraphics[scale=0.52]{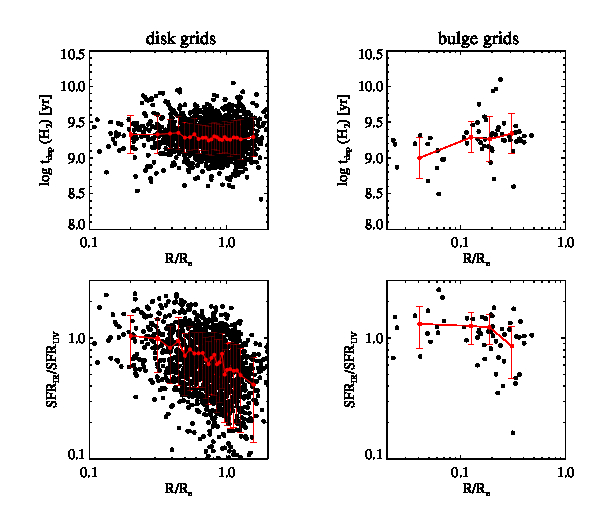}
  \caption{Molecular gas depletion time (upper panel) and the IR/UV ratio (lower panel)
	as a function of radius for disc and bulge grids of HERACLES sample. 
	The radius is normalized to the effective radius of the disc, R$_e$. }
\end{center}
\end{figure*}

\subsubsection{COLD GASS} 						
As in the previous section, we use GALEX FUV and WISE 22$\micron$ images to estimate SFR 
for COLD GASS galaxies within the central 22\arcsec\ region.  
The procedure is the same as mentioned above.
For consistency, we use the same calibration as we apply to HERACLES and ATLAS$^{3D}$ 
sample to convert FUV plus 22$\micron$ into SFR.  

\subsection{Derivation of stellar mass}                                         
Stellar mass is derived using the SED-fitting method in \citet{wan} by which
the stellar population synthesis models of \citet{bru} are fit to SDSS 5-band magnitudes.
The parameter, sSFR, is the SFR divided by the stellar mass.

\section{Results} 														

\subsection{The effect of internal structures on molecular gas depletion time }
\subsubsection{Bulges and discs}											
We plot the depletion time and the ratio of  SFR$_{22\mu m}$ to SFR$_{\rm FUV}$
(IR/UV) as a function of radius for bulge and disc grids in Figure 6. 
Whereas the depletion time of the bulge grids drops towards the center of the galaxy, 
the depletion time of the disc grids remains almost constant as a function of radius. 
In contrast, the IR/UV ratios of the disc and bulge grids decline from 
the center towards the outskirts -- this is not surprising because   
the IR/UV ratio is a robust indicator of dust attenuation, and  the outer 
parts of galaxies are expected to have lower metallicities and dust content. We note that the 
fact that different radial trends are found for depletion time and for IR/UV ratio
make it seem unlikely that variations in CO-to-H$_2$ conversion factor  
are responsible for the drop in the depletion time in the central bulge.              
We note that this result is consistent with results presented  in \citet{ler13} 

\begin{figure} 
\begin{center}
\includegraphics[scale=0.4]{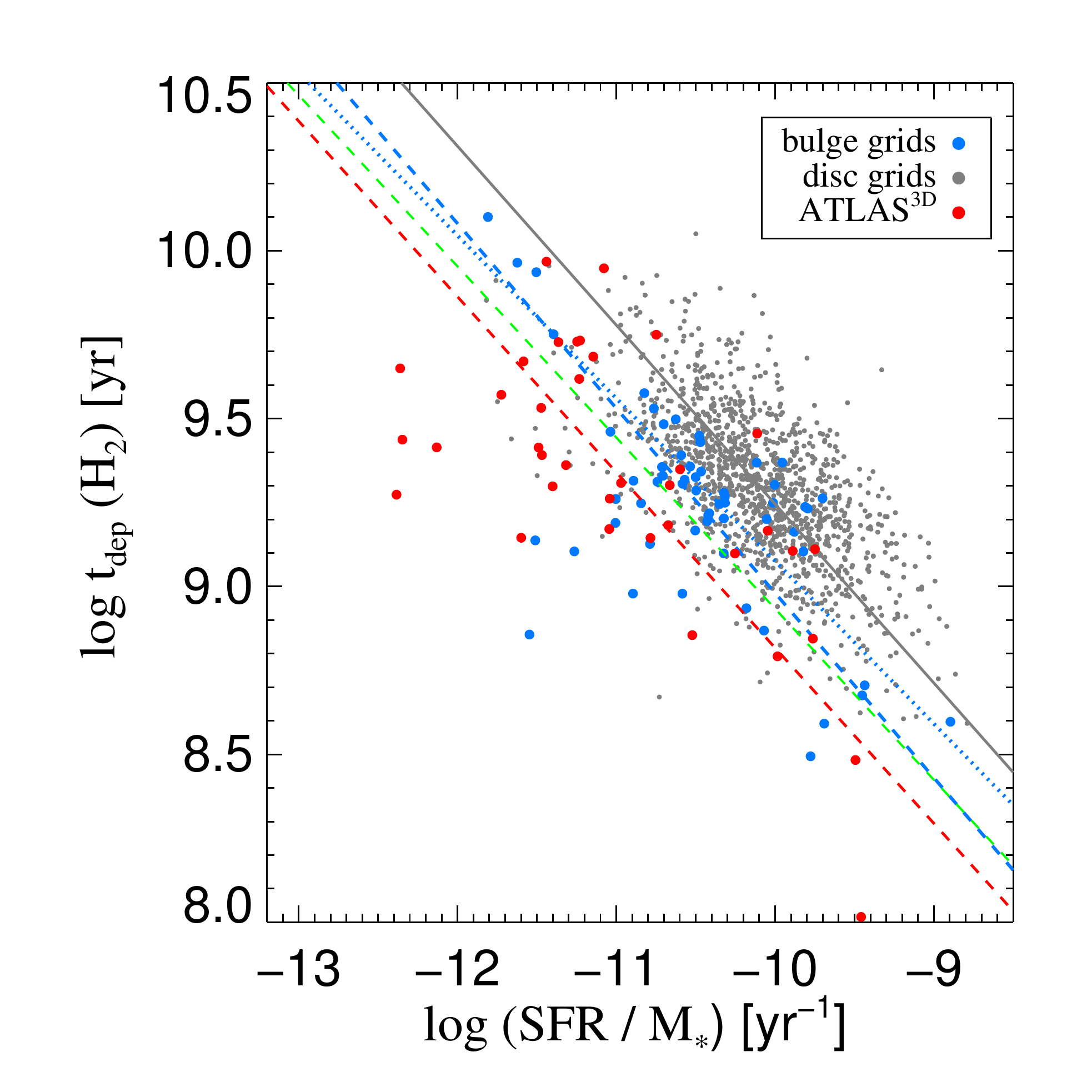}
  \caption{t$_{dep}-$ sSFR relation for the bulge and disc grids of the HERACLES sample,
as well as the   
 ATLAS$^{3D}$ galaxies.  
Grey and blue solid points show grids in the disc and bulge regions of the HERACLES galaxies.  
Red solid points show the ATLAS$^{3D}$ sample. OLS bisector fits to the disc, bulge and  
ATLAS$^{3D}$ data points are shown as grey, blue, and red dashed lines.
The blue dotted line shows the OLS bisector fit to the bulge grids, excluding those 
those with R $<$ 0.1$R_e$. The green dashed line is the OLS bisector fit to the HERACLES bulge 
data points plus the results from the ATLAS$^{3D}$ sample.}
\end{center}
\end{figure}

In Figure 7, we investigate the t$_{dep}-$sSFR relation for bulges (blue solid points) 
and discs (grey solid points) regions.  We also 
plot the ATLAS$^{3D}$ bulge-dominated  sample for comparison (red solid points).   
We perform OLS-bisector fits to the data and plot the fitting results in Figure 7 as well.
There are two obvious results: First, the scatter in depletion time is larger in bulges 
than in discs (0.33 dex in $\log t_{dep}$, compared to 0.22 for discs). 
Excluding  bulge grids with R $<$ 0.1 R$_e$, which may be influenced by AGN , makes 
no difference to this conclusion.
Second, in the bulge, t$_{dep}$ is systematically shifted to smaller values at  
a given sSFR compared to the disc grids.
We note that there are some bulge grids whose depletion times lie  above the 
grey line derived for disc grids in Figure 7. 
We check the location of these grids and find that they are generally in the outer
region of the bulge.

\subsubsection{Spiral arms, bars and rings}				
In Figure 8, we plot the t$_{dep}-$sSFR relation for galaxies
with  grand-design arms,  bars and rings. Magenta points denote the grids located within
the region of the targeted structure, such as the grand-design arm, bar or ring; 
blue points denote other grids in the same galaxies.  Grids from other galaxies,
i.e. those galaxies that do not contain the designated structure,  
are shown in grey.

\begin{figure*} 
\begin{center}
\includegraphics[width=.33\textwidth]{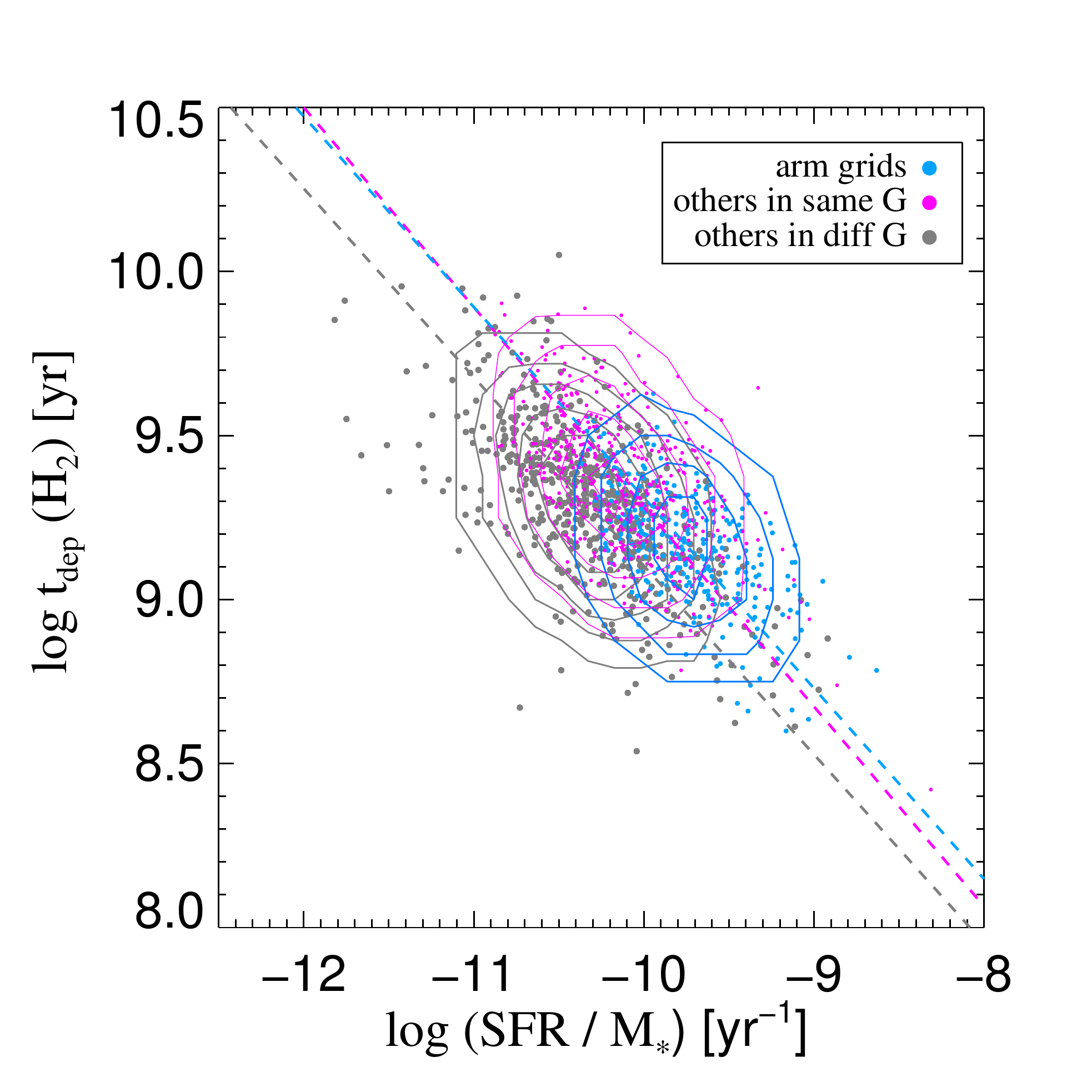}
\includegraphics[width=.33\textwidth]{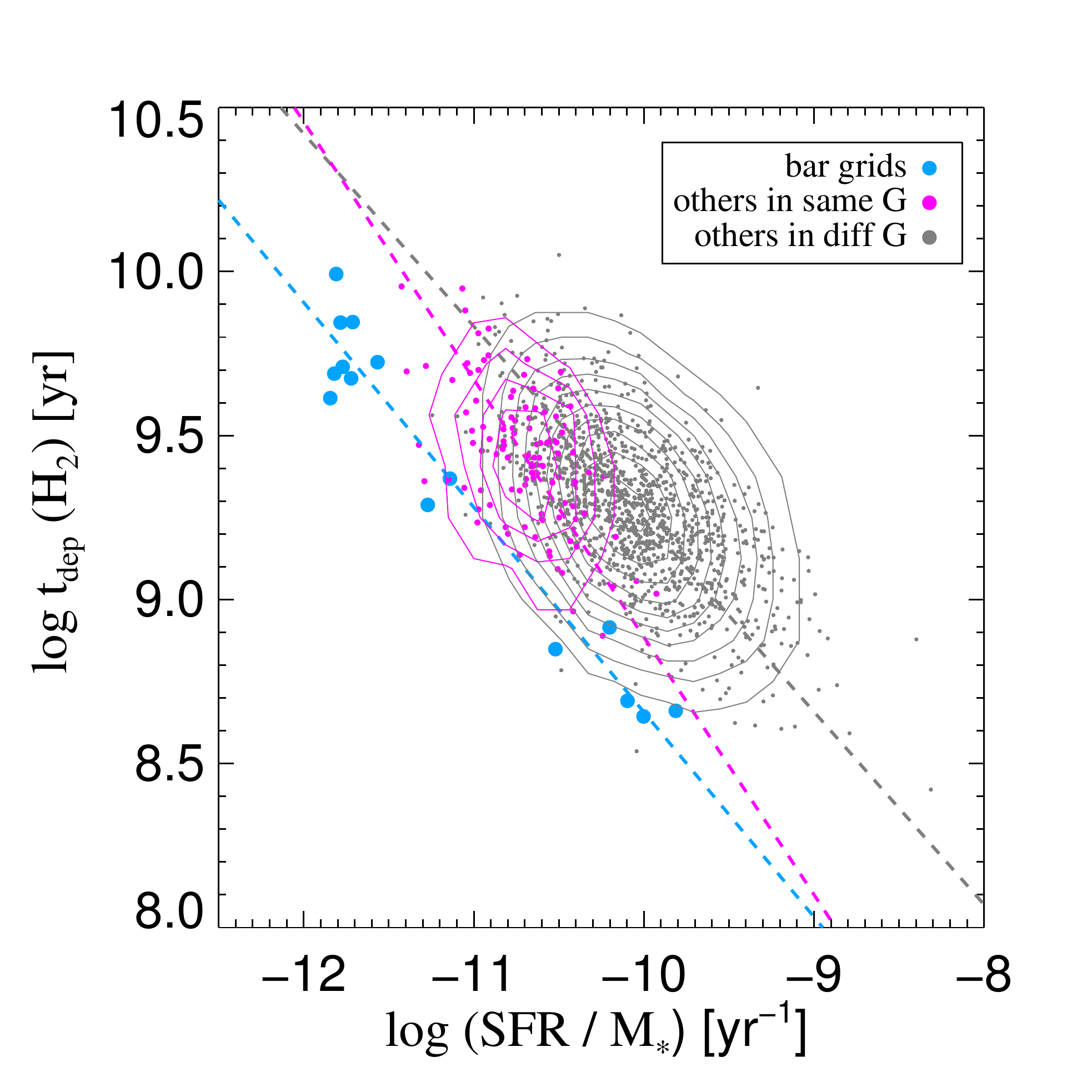}
\includegraphics[width=.33\textwidth]{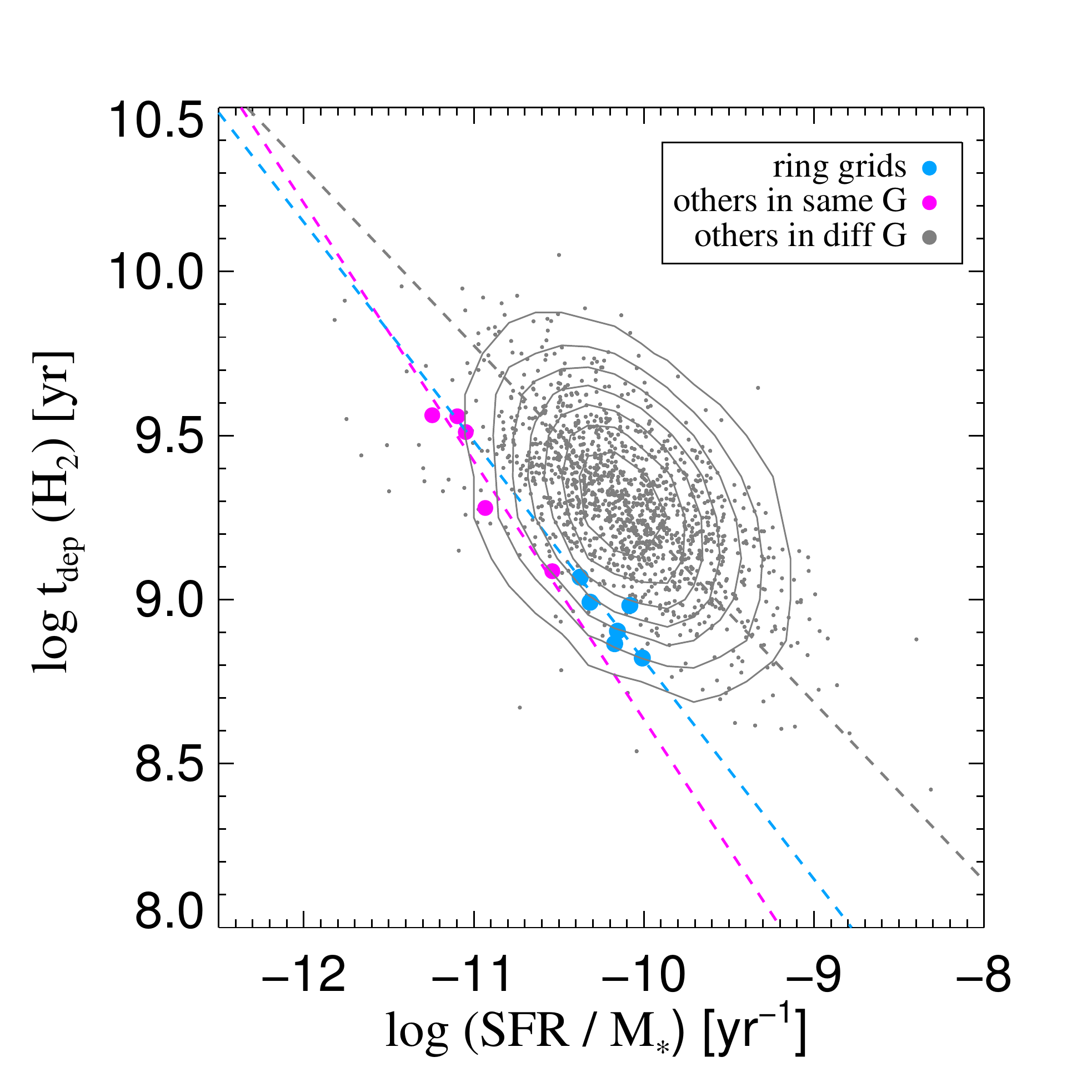}
  \caption{Distribution in the t$_{dep}-$ sSFR plane of grids in galaxies with spiral arms
(left), in galaxies with bars (middle) and in galaxies with rings (right). Results
for grids within the arm, bar or ring regions are coloured in blue.
Results for grids in these galaxies outside the arms,bars or rings are coloured
in magenta. Results for grids in galaxies without arms, bars or rings are coloured in grey.
OLS-bisector fits to the data points in these different groups are shown
with the same color.}
\end{center}
\end{figure*}

\begin{figure*} 
\begin{center}
  \includegraphics[width=.33\textwidth]{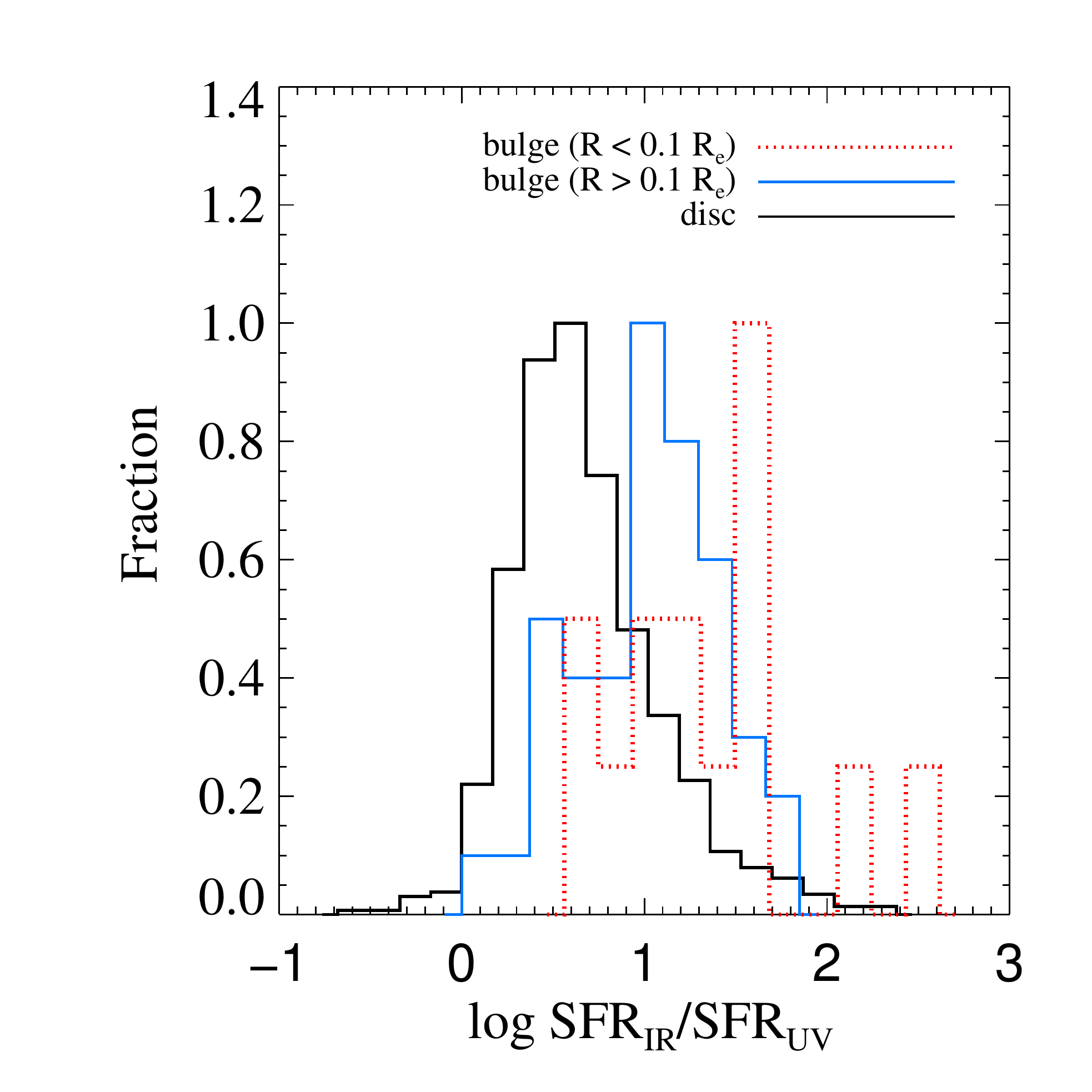}
  \includegraphics[width=.33\textwidth]{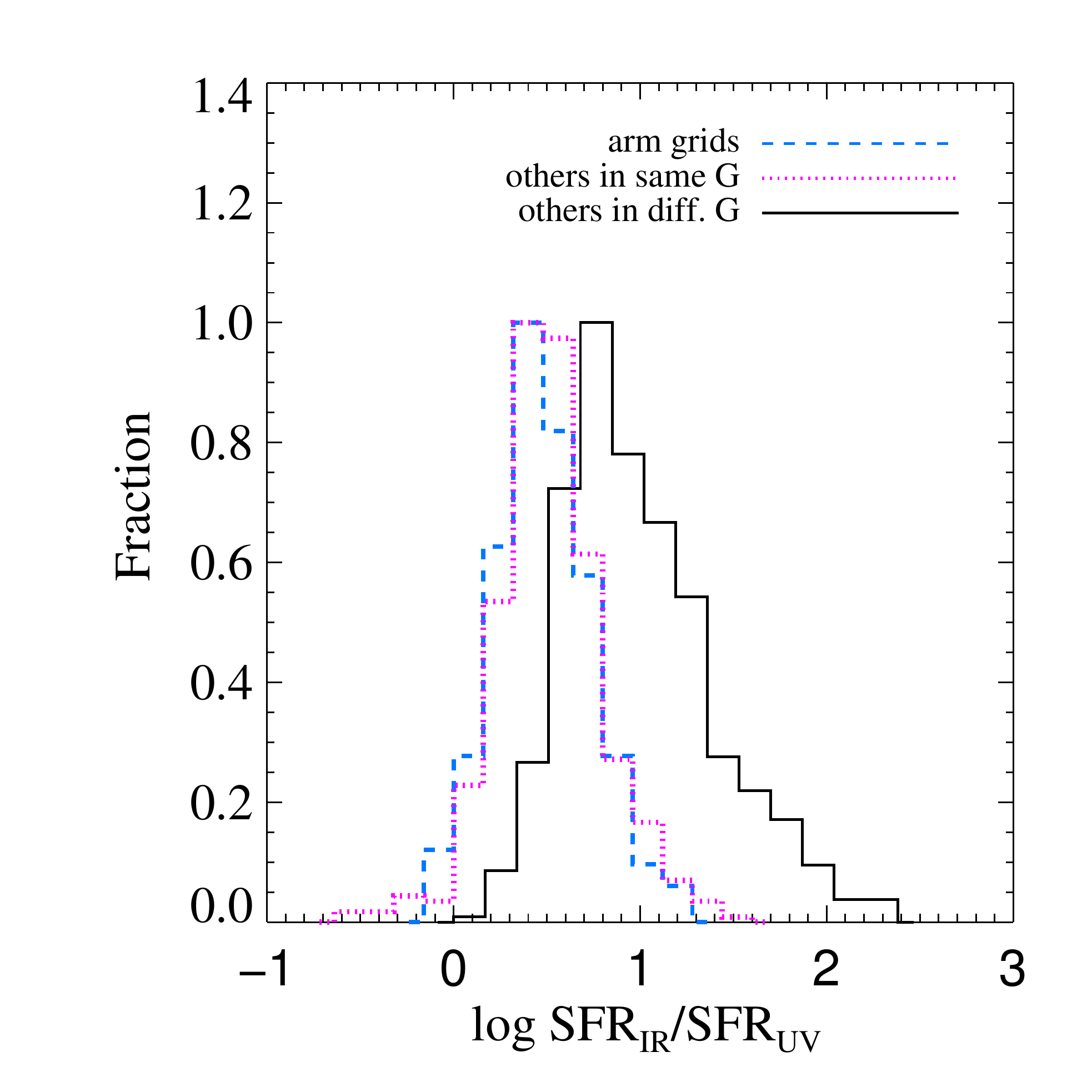}
  \includegraphics[width=.33\textwidth]{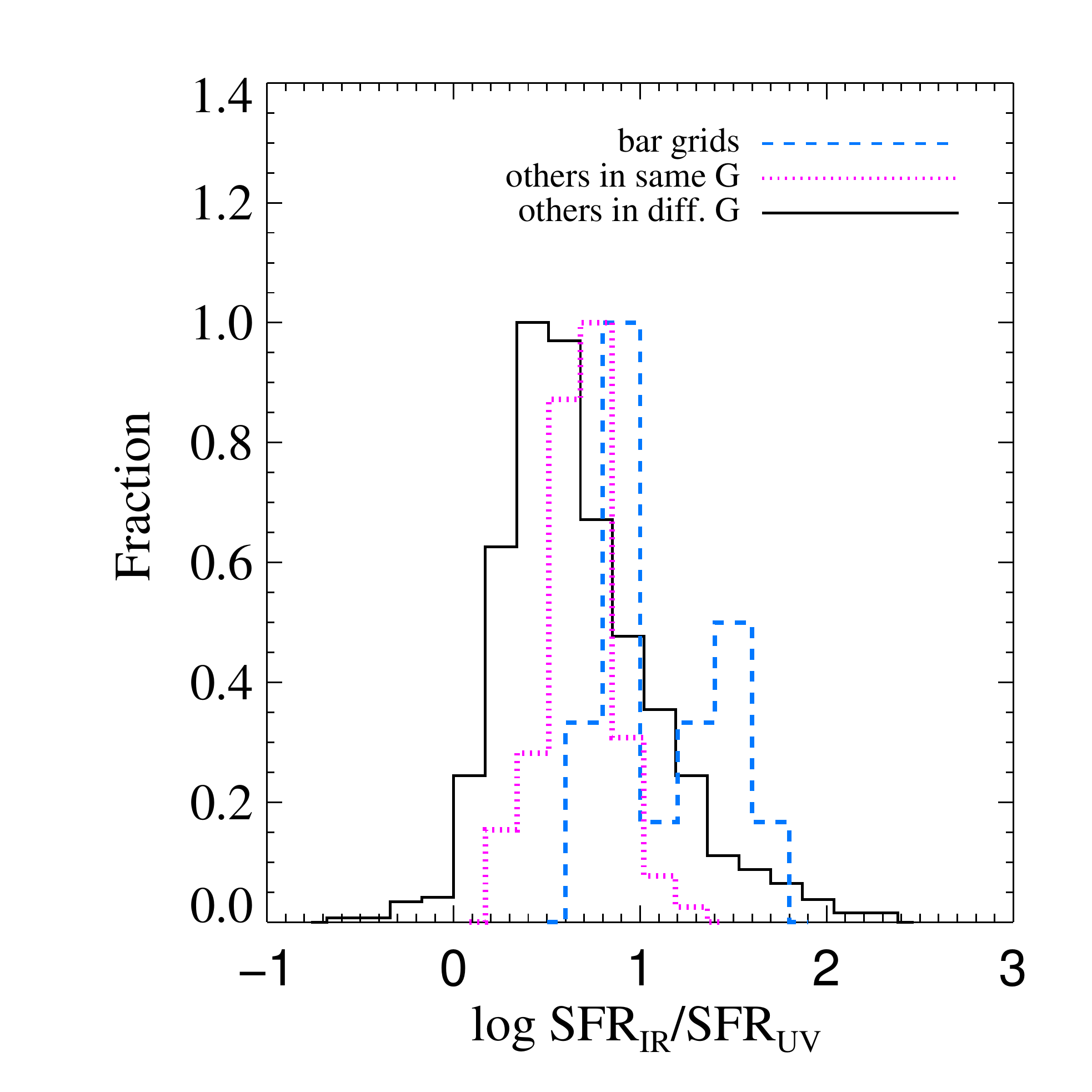}
  \includegraphics[width=.33\textwidth]{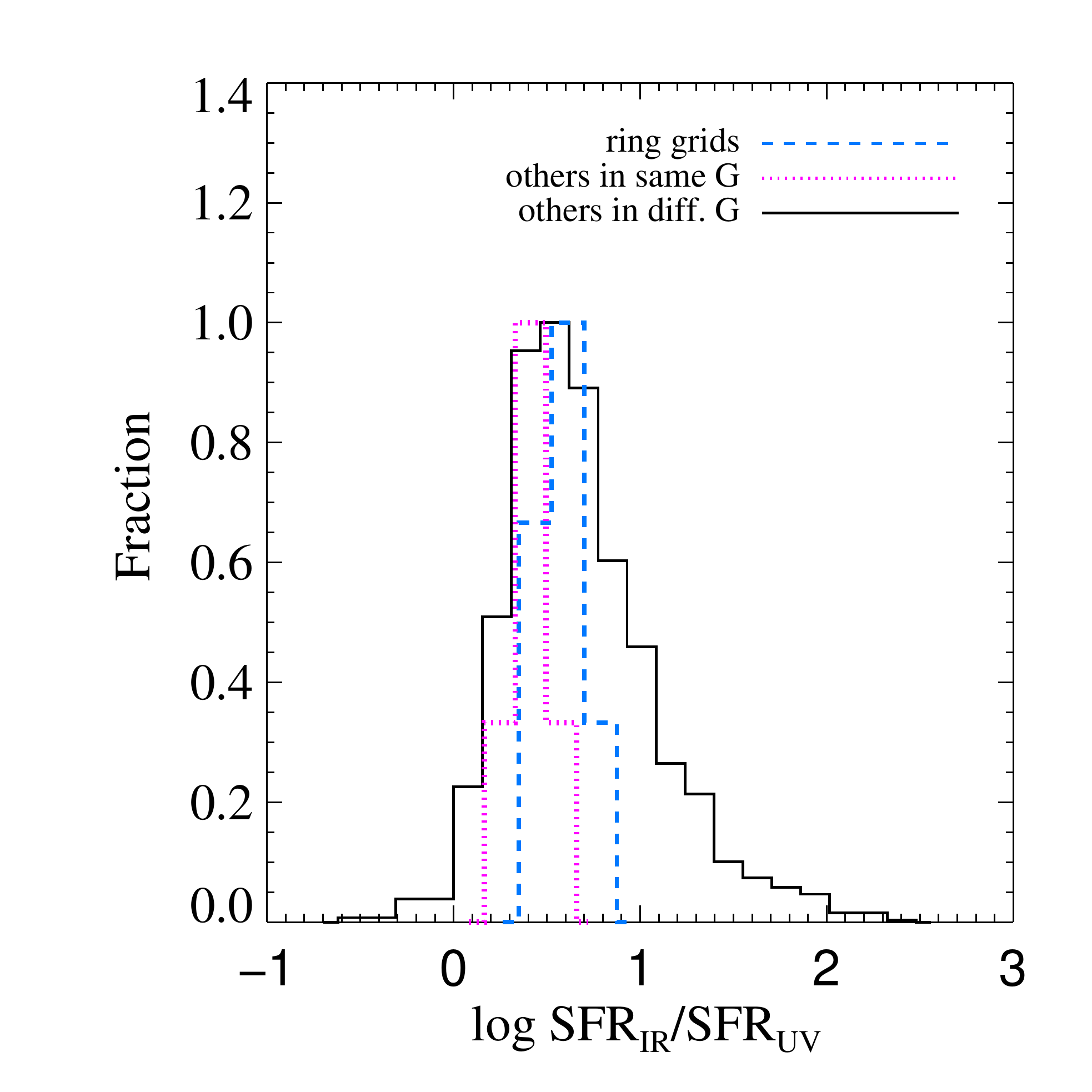}
    \caption{Distribution of the  ratio of  SFR$_{22\mu m}$ to SFR$_{\rm FUV}$. 
        In the upper left panel,
	  red dashed and blue histograms show results for grids in the bulge regions with
	  R $< 0.1R_e$ and R $> 0.1R_e$ respectively. The  black histogram is for the grids 
	  in the disc regions. In the other panels, blue histograms are for grids within 
	  the structure (arm or bar or ring); magenta histograms are for 
	  the grids outside the structure; black histograms are for the grids from 
	  the other galaxies in the HERACLES sample. Upper right panel: the spiral-arm galaxies. 
	  Lower left and right panels: the barred and ring galaxies.}
\end{center}
\end{figure*}

\begin{figure*} 
\begin{center}
  \includegraphics[width=.33\textwidth]{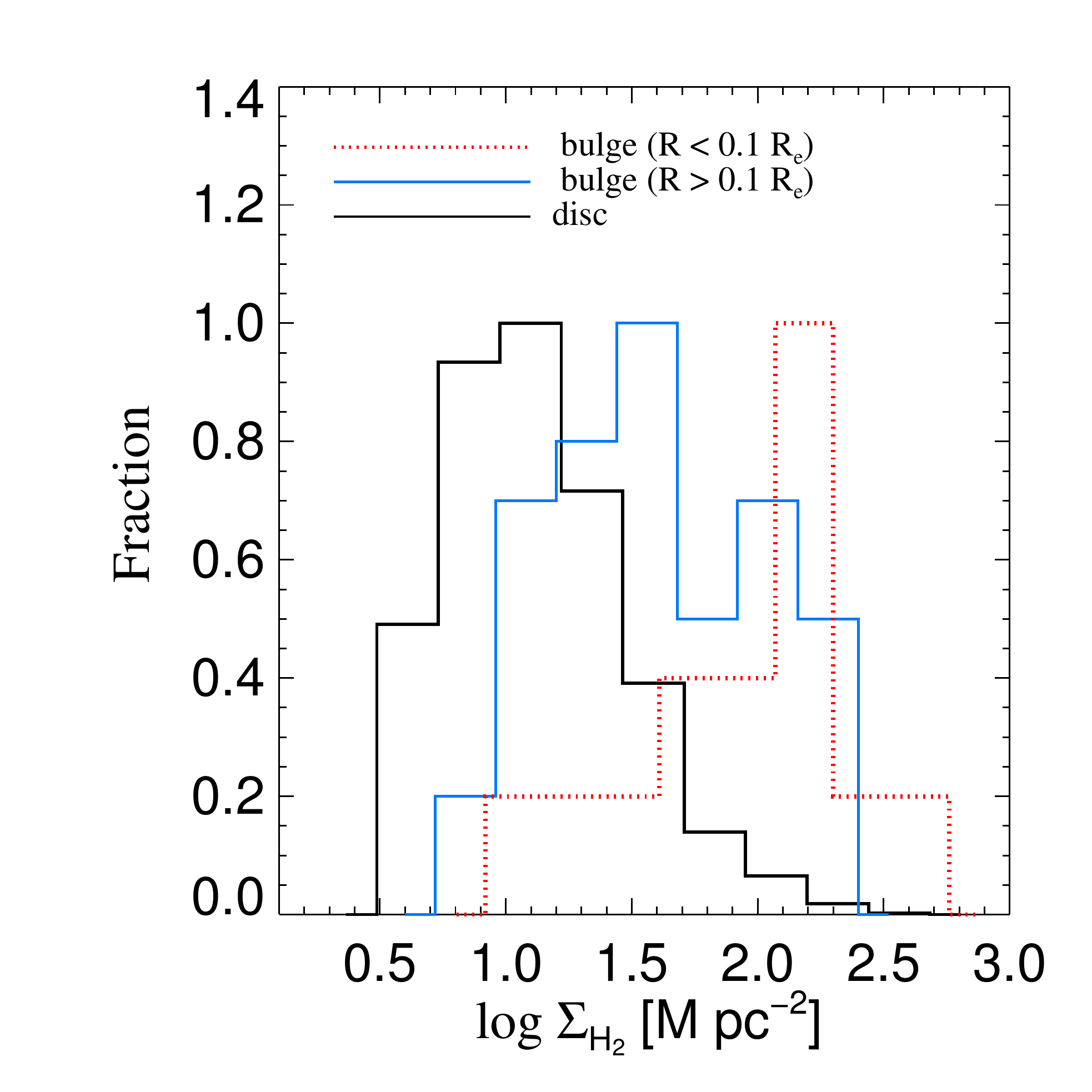}
  \includegraphics[width=.33\textwidth]{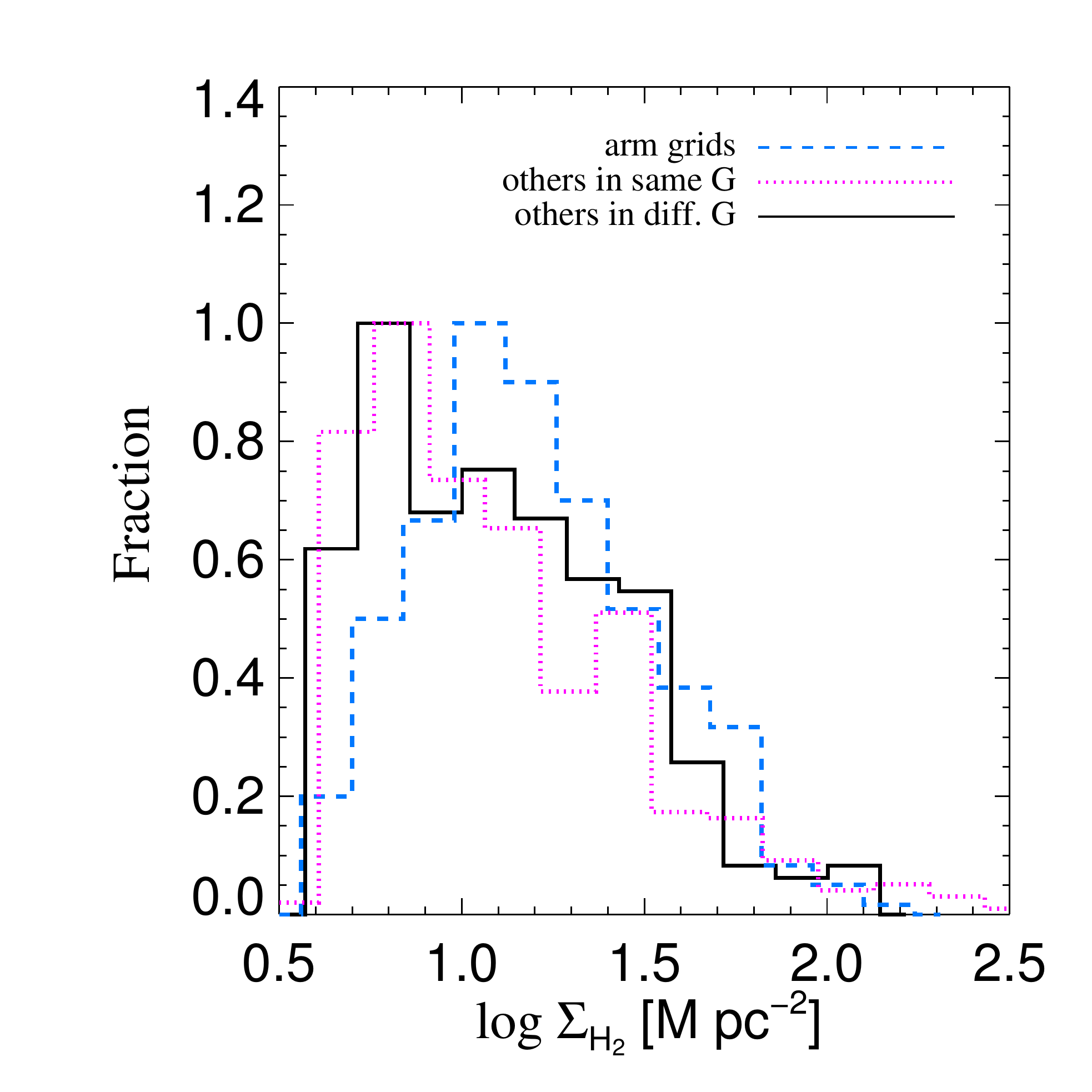}
  \includegraphics[width=.33\textwidth]{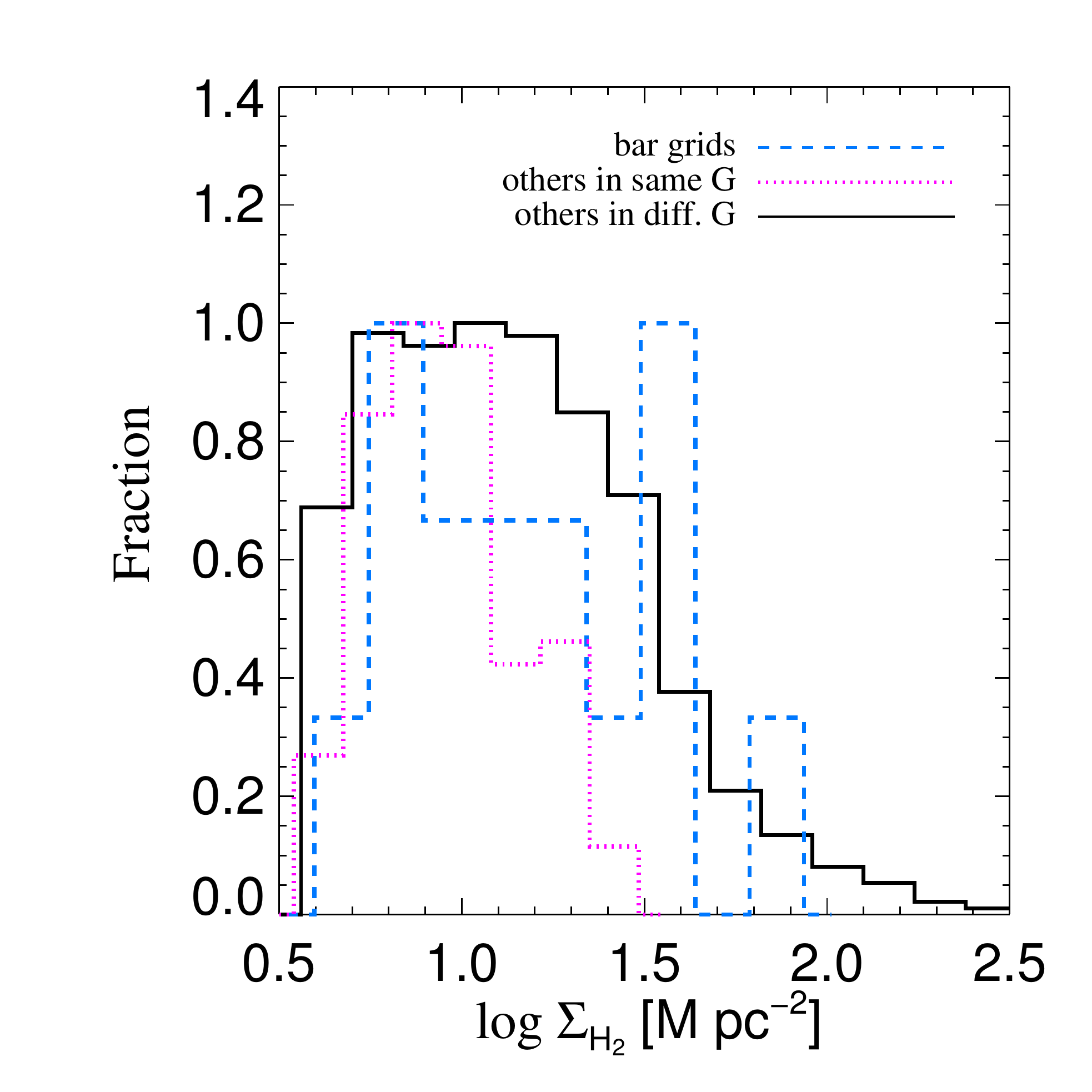}
  \includegraphics[width=.33\textwidth]{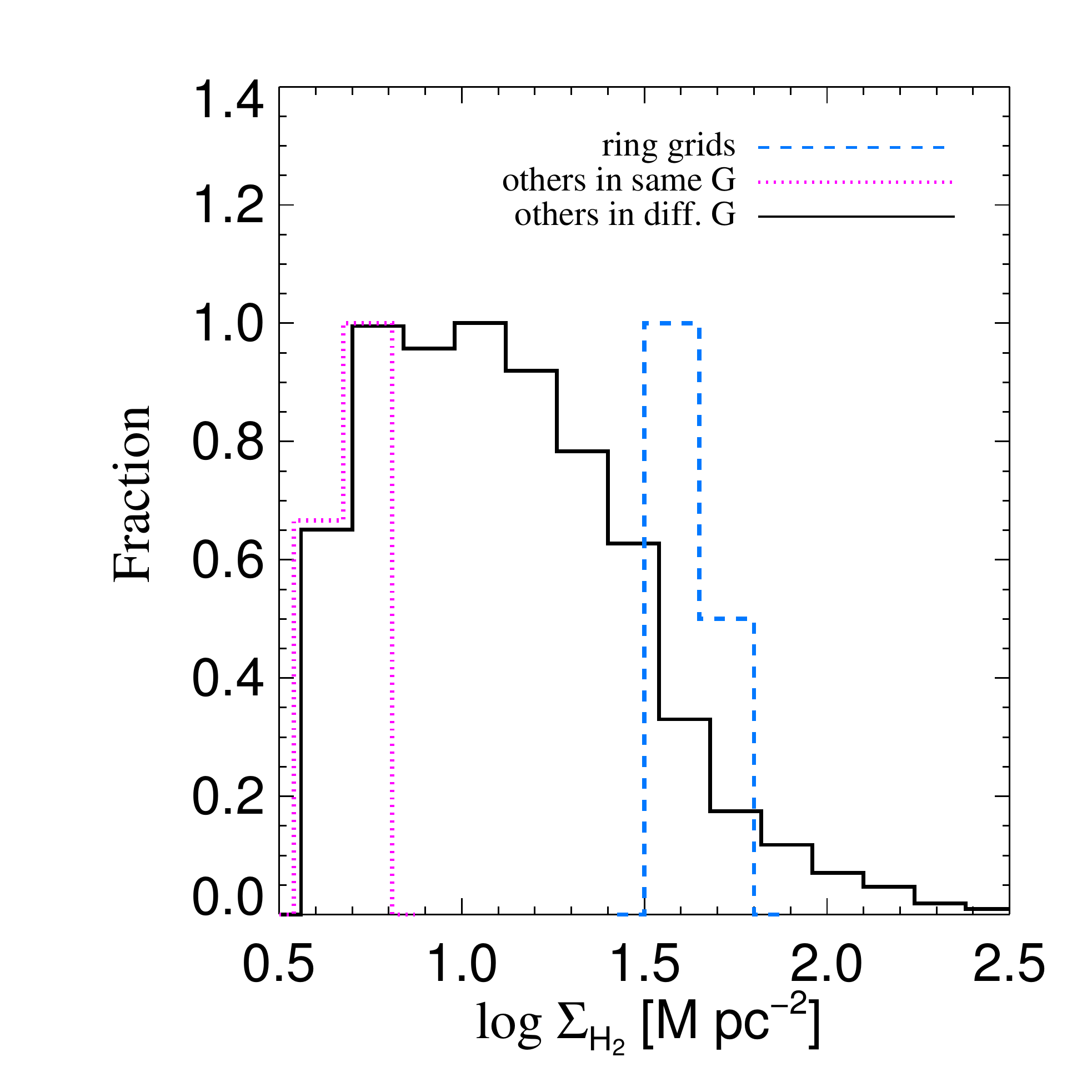}
    \caption{ Distribution of the molecular gas surface densities. 
	  Conventions and lines are as in Figure 9.}
\end{center}
\end{figure*}

\begin{figure*} 
\begin{center}
  \includegraphics[width=.33\textwidth]{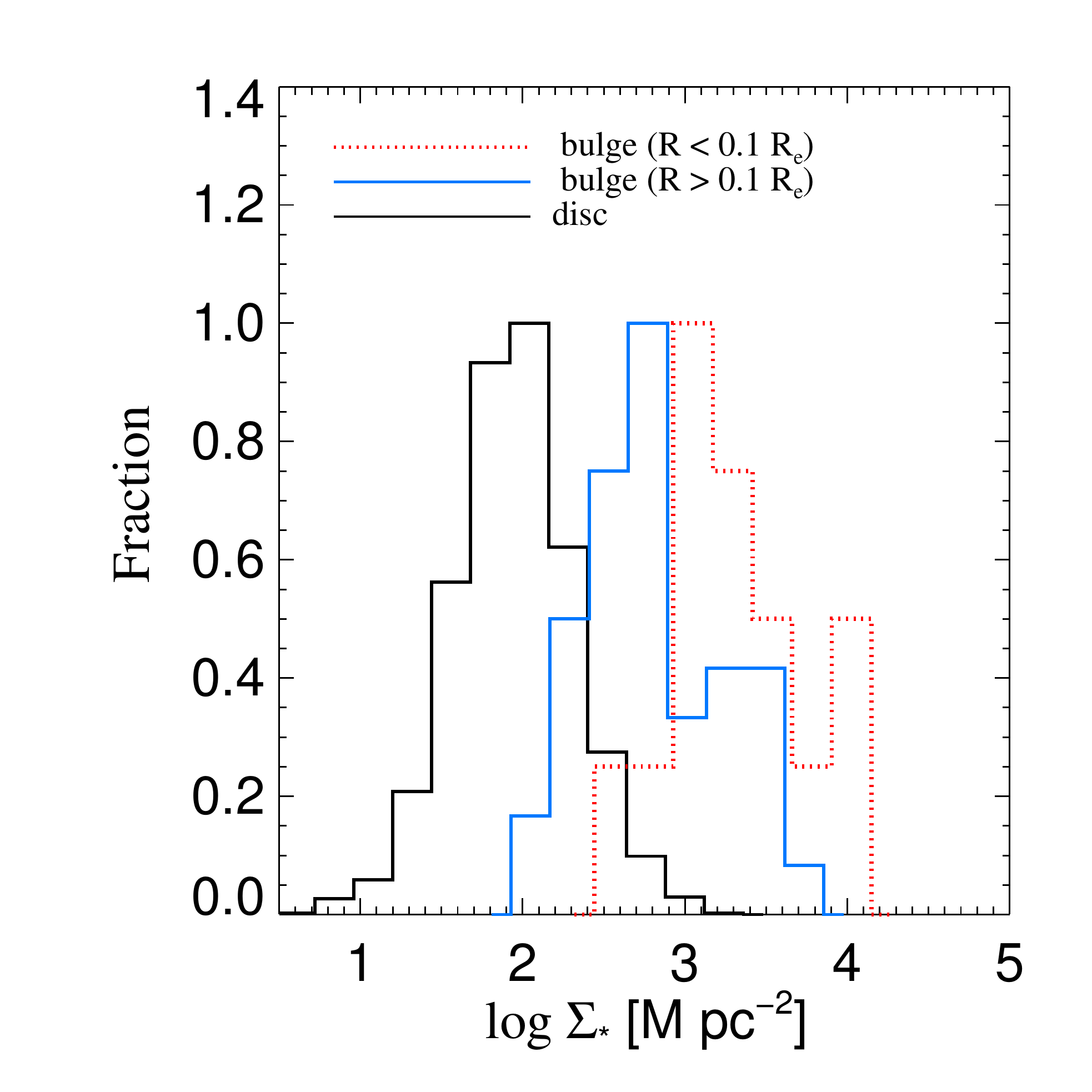}
  \includegraphics[width=.33\textwidth]{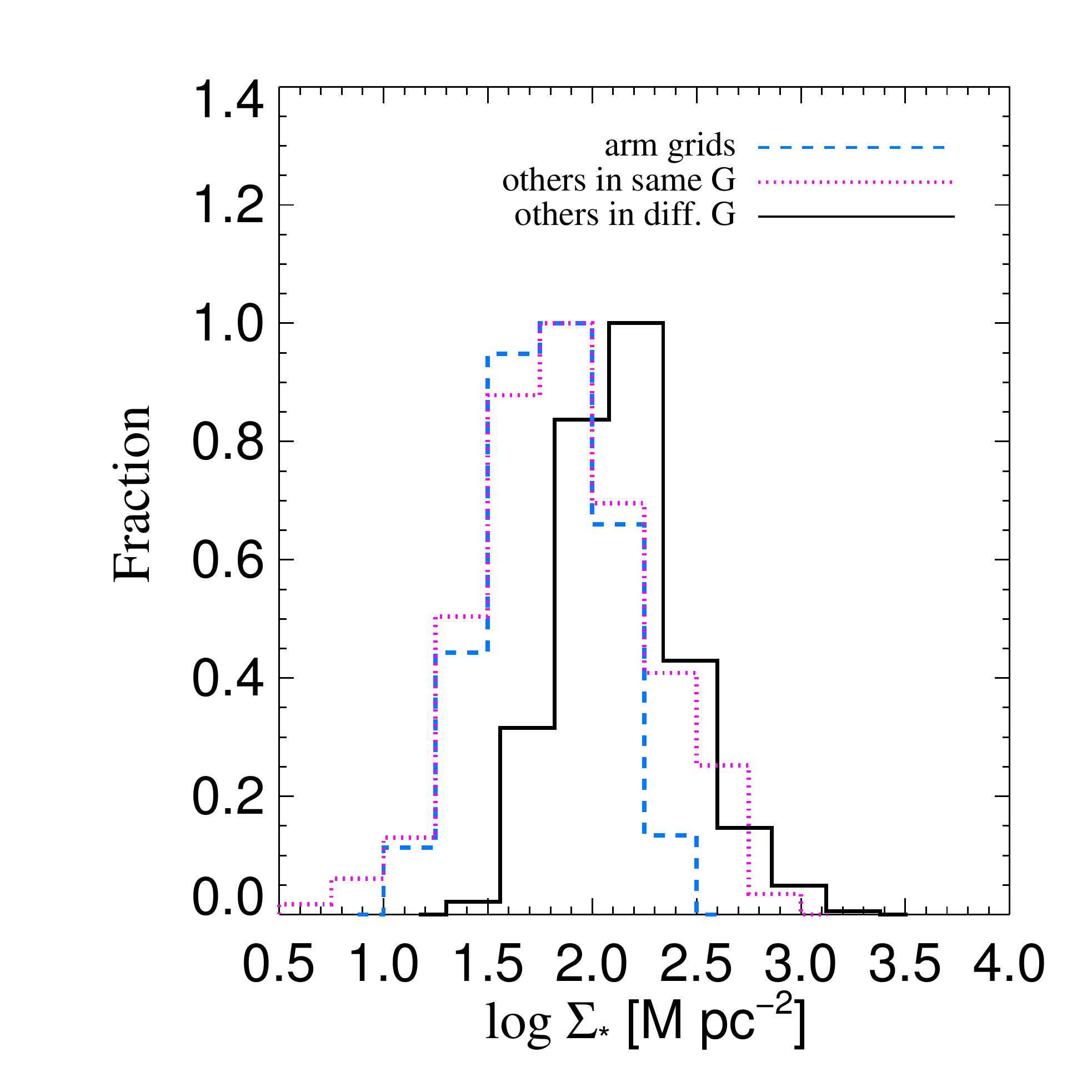}
  \includegraphics[width=.33\textwidth]{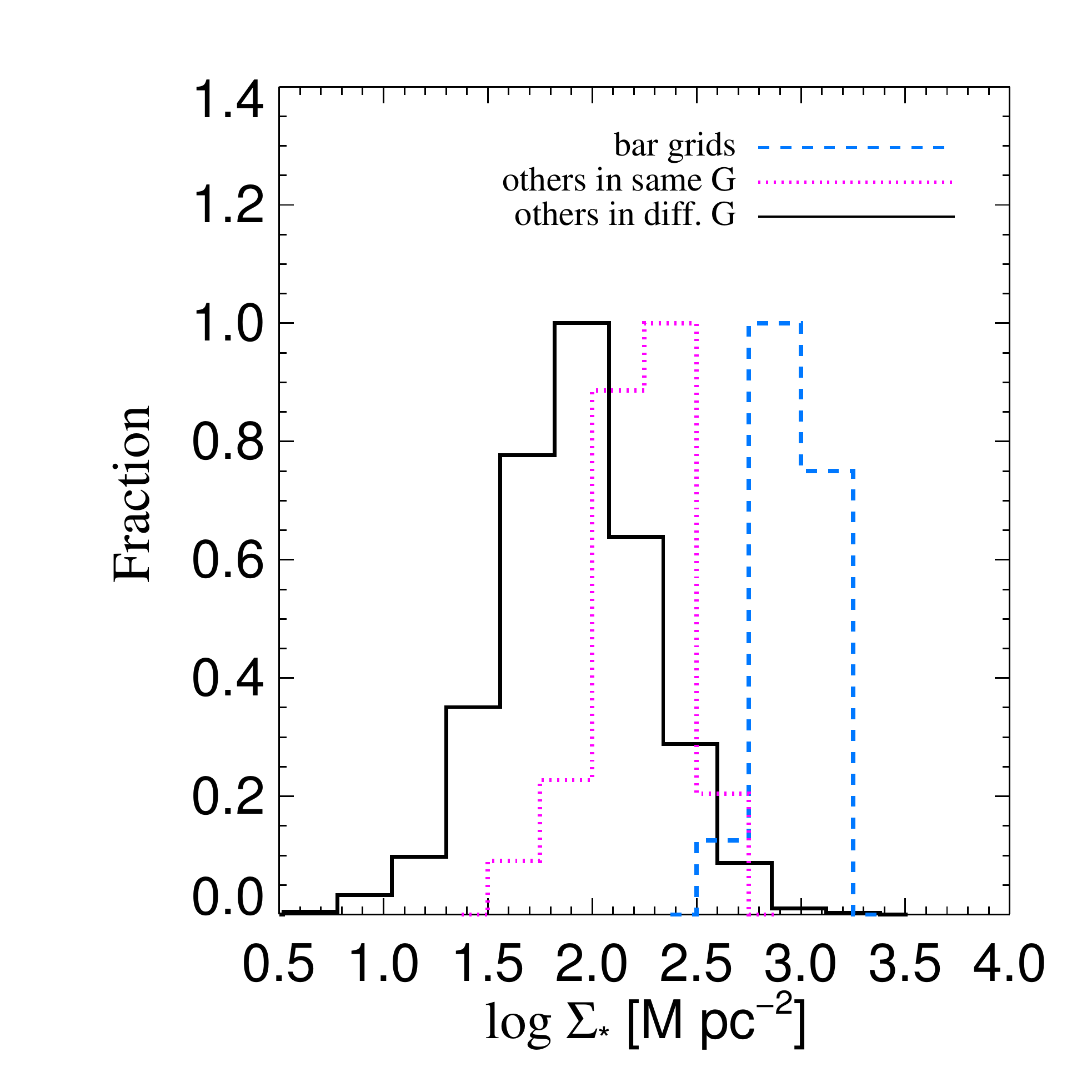}
  \includegraphics[width=.33\textwidth]{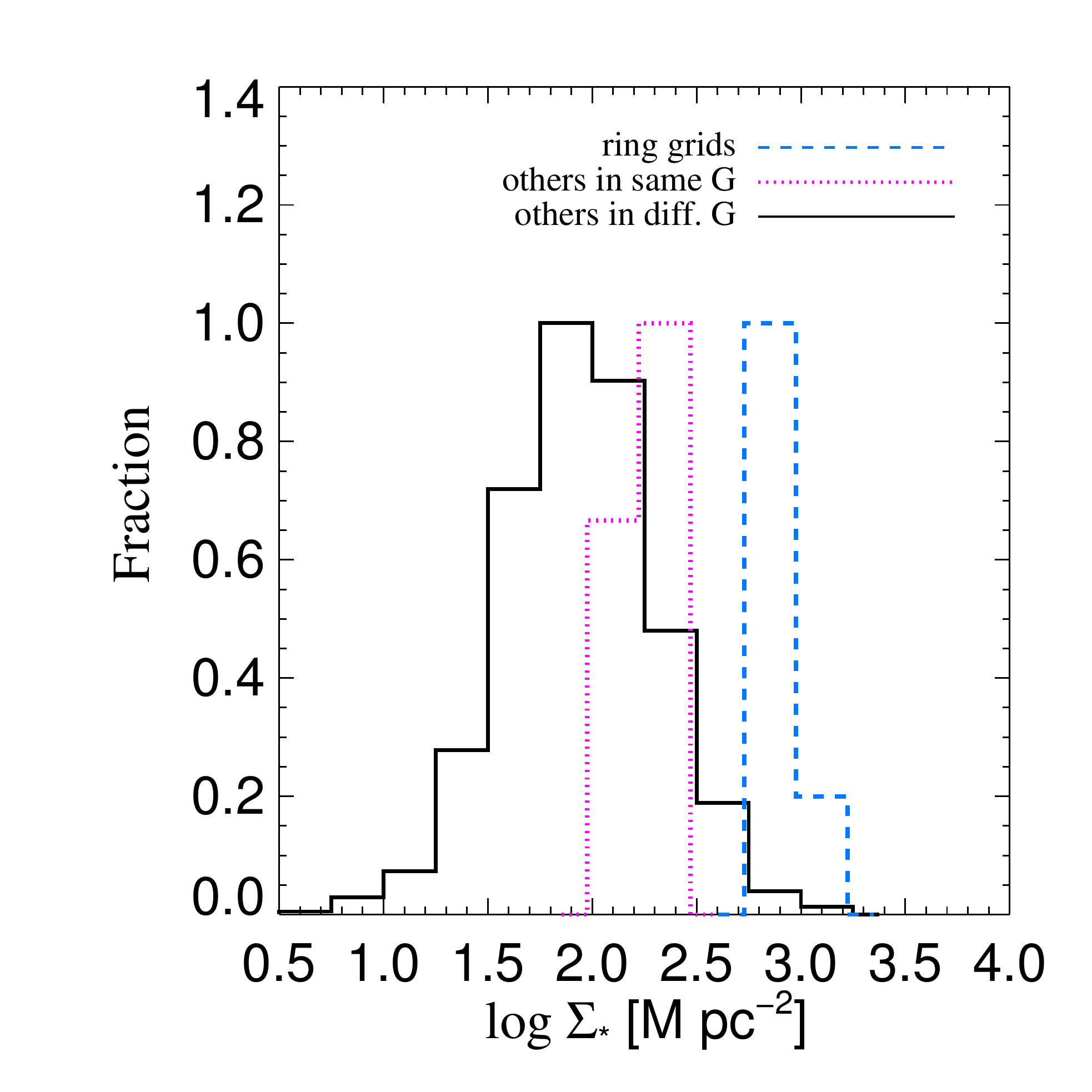}
    \caption{ Distribution of the stellar surface densities.
	  Conventions and lines are as in Figure 9.}
\end{center}
\end{figure*}

The sSFRs in the interstellar medium (ISM) of galaxies with grand-design arms are higher than
in galaxies without arms, and the sSFRs within the arms themselves are the highest.
This is expected, because gas is compressed by spiral density waves and
star formation rate scales with gas surface density. Interestingly, t$_{dep}$ is shifted 
to higher values at a given sSFR in galaxies with spiral arms. 
We apply OLS-bisector fits to the spiral arm data and the results are shown  as dashed lines
in Figure 8. Indeed, the grids in the regions with (blue dashed line) or without  
(magenta dashed line) arms are systematically shifted to  longer t$_{dep}$ at fixed sSFR
compared to the  grey dashed line, which is for galaxies without spiral structure.  

In barred galaxies, we find the opposite trend.  
The sSFRs in the interstellar medium of galaxies with bars (magenta points in the middle panel)
are much lower than in galaxies without bars. Bars are long-lived structures.
Although they can act to channel gas towards the center of a galaxy and fuel
starbursts, this is a transient phenomenon and the majority of present-day barred galaxies 
are relatively quiescent.
The OLS-bisector fit to the grids in barred galaxies indicates that 
t$_{dep}$ is shortened at fixed sSFR compared to grids from galaxies without bars. 
This is exactly the opposite to what was found for galaxies with spiral arms.
There is only one galaxy with a ring in HERACLES sample, so we only have a few grids  
available for analysis. Based on the limited data, we conclude that ringed galaxies
follow the same trends as barred galaxies.

\subsection {Linking the results together}

So far, we have analyzed molecular gas depletion time variations in terms of
location within or outside specific 
structures of the galaxy, such as bulges, bars, arms or rings. 
We now ask whether all of these variations can be linked to a physical parameter
that varies continuously within galaxies. For example,
Daddi et al. (2010) and Genzel et al. (2010) proposed that the combination of
high gas density and short dynamical timescale could explain the
short global  molecular gas consumption timescales in strongly  starbursting galaxies. 
The parameters that we examine in this section are, 1) IR/UV ratio as a proxy for
dust content, 2) H$_2$ surface density, 3) stellar surface density.

{\bf {\em IR/UV ratio}}
In Figure 9, we plot the distributions of IR/UV ratios for inner bulge, outer bulge
and disk grids in the top left panel. In the top right panel,
we plot distributions  for spiral arm, inter-arm and control grids. The bottom two
panels are the same as the top right panel, except that results are shown for
bars and rings.

As we have seen previously, IR/UV ratios are highest in the inner regions of bulges,
from which we infer that these are the dustiest regions of nearby galaxies.
The high dust content in the centers of bulges means that it is unlikely that
we have under-estimated the H$_2$ content on the galaxy by using the Galactic
CO-to-H$_2$ conversion factor. The short depletion times observed in central
bulges are thus not likely to be an artifact of wrongly estimated molecular
gas masses.
The least dusty regions of the ISM are found within galaxies with grand-design
spiral arms. By a similar argument, we infer that that higher molecular
gas depletion times observed in these regions are not likely to be
an artifact of conversion factor variations either. 

Grids within bars in barred galaxies have similar
IR/UV ratio as the outer parts of galactic bulges, while those outside bars
have IR/UV ratios similar to the general disc population. Grid within rings
appear less dusty than grids within bars, but we caution that our sample
consistent of only one ringed system.  

In summary, we deem it unlikely that variations in CO-to-H$_2$ conversion factor 
are responsible for the different t$_{dep}-$sSFR relations among grid cells lying 
within  different structures.

{\bf{\em H$_2$ surface density}}
Figure 10 is the same as Figure 9, except that distributions of H$_2$ surface density are 
shown. If we exclude the single ringed galaxy, the only significant shift in H$_2$ surface
density occurs within bulges, which have molecular gas densities up to a factor of
30 higher than discs, particularly in their inner regions. The enhancement in gas density 
in the grids with spiral arms is only about 50\%, and grids that lie outside
spiral arms have H$_2$ densities that are identical to grids with galaxies without arms.
Gas density is enhanced more strongly within bars, but still not as strongly as in bulges. 

In conclusion, although high gas densities may be a promising avenue for explaining the short
depletion time in bulges and starburst systems, it does not appear to vary in a systematic way 
in other environments such as spiral arms and bars, so as to explain other observed trends.
We will demonstrate this in more detail later in this section.

{\bf {\em Stellar surface density}}
In Figure 11, we plot the distributions of stellar surface density for the grids
within different structures. In this case, there is a very clear separation between
grid cells in bulges, bars and rings, which peak at   
log $\Sigma_{*}\sim3$ M$_{\sun}$ pc$^{-2}$ while  grids in 
the spiral-arm galaxies peak at log $\Sigma_{*}\sim1.5-2$ M$_{\sun}$ pc$^{-2}$.
Grids in the inner regions of bulges, where $t_{dep}$ is shortest, are shifted to
even higher $\Sigma_{*}$ than the outer bulges. Spiral arm and inter-arm  grids also
peak at lower $\Sigma_{*}$  values than grids from galaxies without spiral arms.
In summary, variations in the  $\Sigma_{*}$ distributions between
different galaxy structures mirror  the  variations
in the $t_{dep}$-sSFR relations described in the previous section extremely well.

We now fit a linear relation to the t$_{dep}-$sSFR relation for all grid data points
and plot the residuals against  
$\Sigma_{*}$, $\Sigma_{\rm H_2}$ and a third quantity, the interstellar
pressure, in Figure 12.
Following \citet{ler08}, the pressure, P$_h$, is expressed as   
\begin{equation}
  \rmn{P_{h}} \approx {\rmn{\pi}\over \rmn{2}} G\Sigma_{gas} (\Sigma_{gas}+ {\rmn{\sigma_{g}}\over \rmn{\sigma_{*,z}}} \Sigma_{*}), 
\end{equation}
where $\sigma_g$ and $\sigma_{*,z}$ are velocity dispersions of gas and stars, and 
$\Sigma_{gas}$ and $\Sigma_{*}$ are the gas and stellar surface densities.
Data from ATLAS$^{3D}$ and COLD GASS are also added in this figure.

Figure 12 shows that  $\Sigma_{*}$ is the only quantity that clearly
correlates with  $\Delta t_{dep}$. Grids in bulges, bars, and rings and the 
ATLAS$^{3D}$ sample clearly separate from the spiral arm, inter-arm and disc grids in this plot.  
We can also understand the reason why we did not observe a correlation 
between $\Delta t_{dep}$ and $\Sigma_{*}$ for the COLD GASS sample in \citet{hua}
: the dynamic range in $\Sigma_{*}$ of the COLD GASS sample is too small.
The integrated stellar surface densities of the COLD GASS sample span only 
1 dex on the x-axis.

\begin{figure} 
\begin{center}
 \subfigure{
    \begin{minipage}[b]{0.53\textwidth}
	   \vspace{-0.1\textwidth}\vspace{0.15in} 
	   \includegraphics[width=0.8\textwidth]{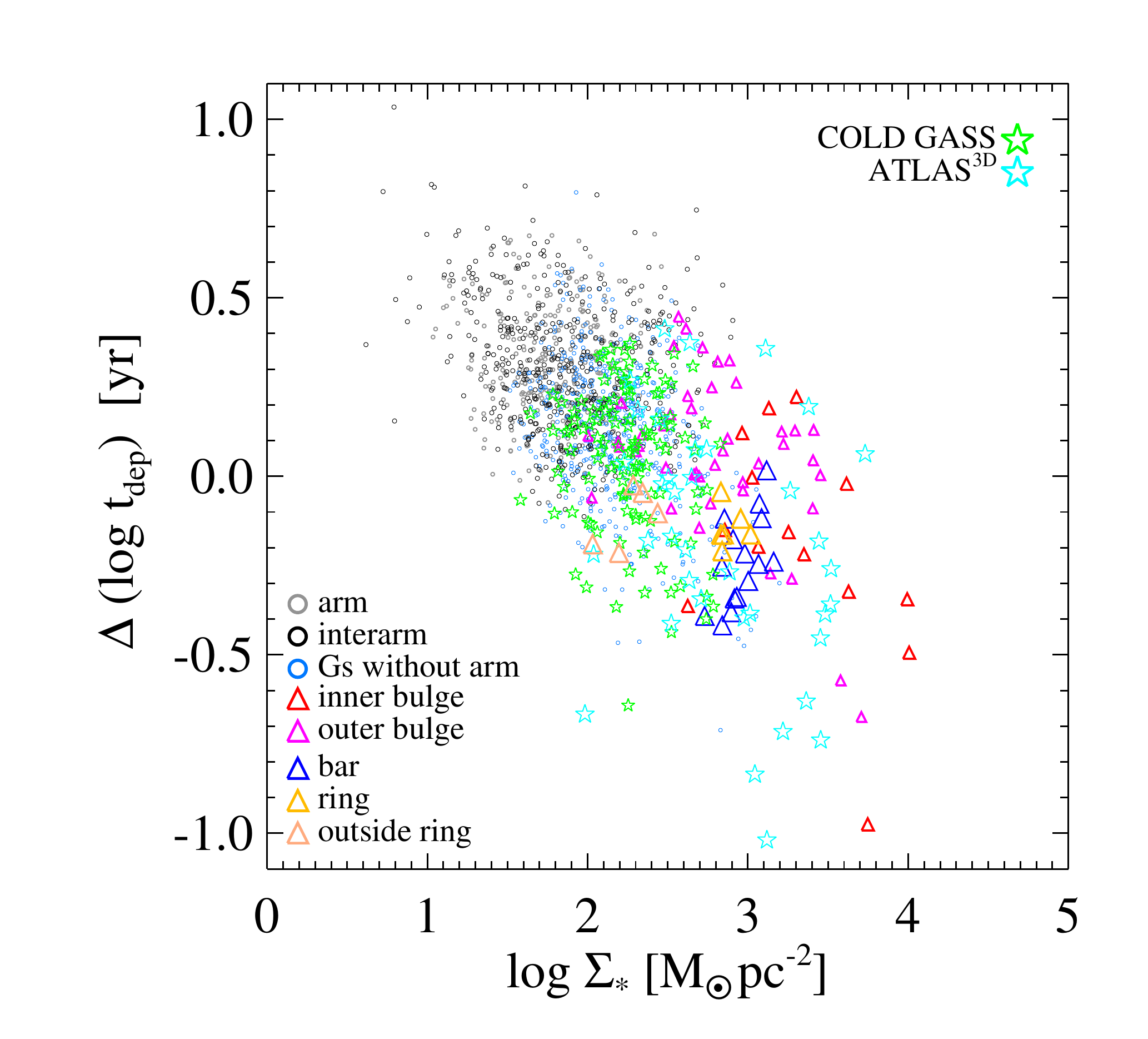}
	 \end{minipage} }
 \subfigure{
    \begin{minipage}[b]{0.53\textwidth}
	   \vspace{-0.1\textwidth}\vspace{0.15in} 
	   \includegraphics[width=0.8\textwidth]{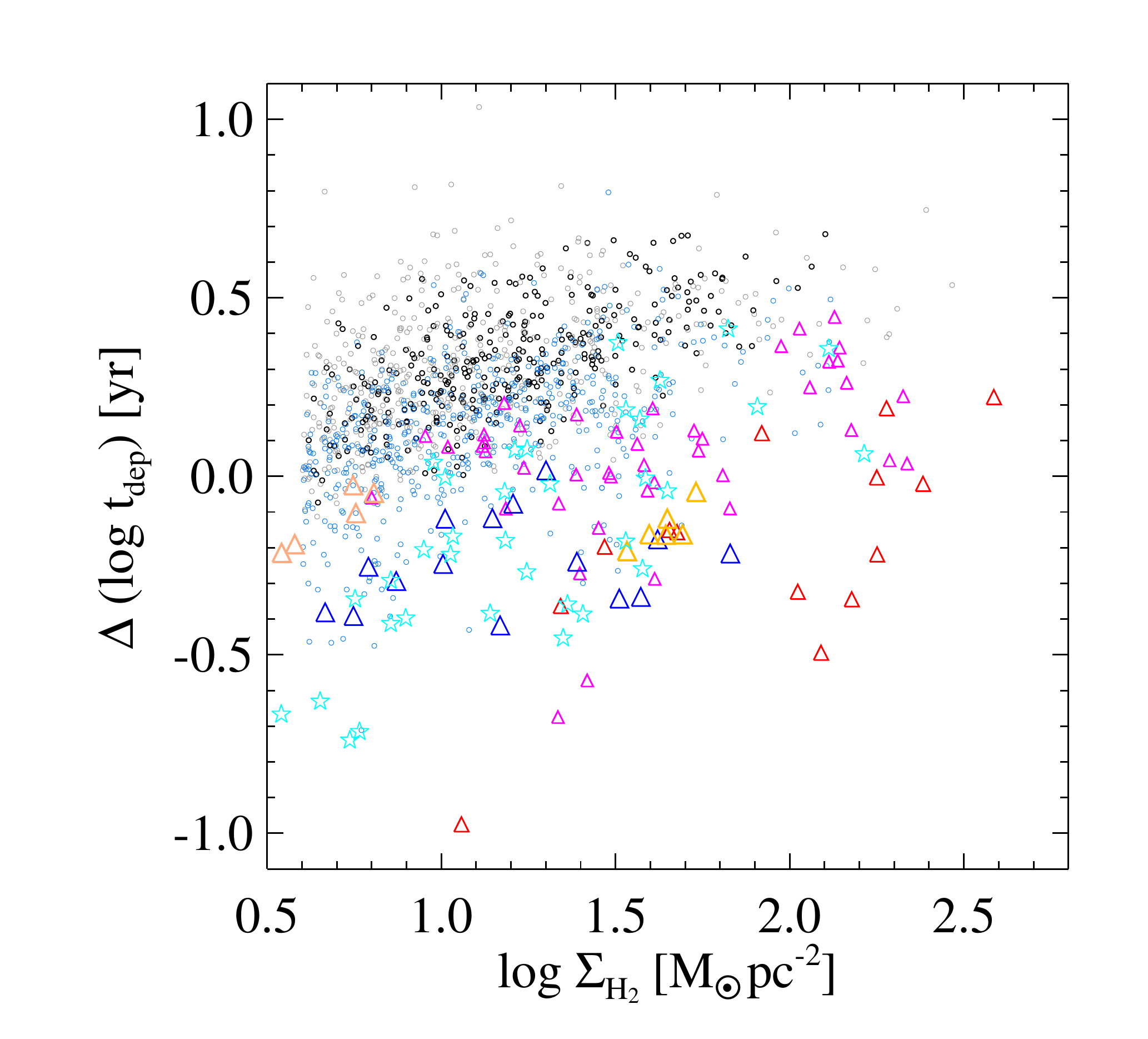}
	 \end{minipage} }
 \subfigure{
 	\begin{minipage}[b]{0.53\textwidth}
	   \vspace{-0.1\textwidth}\vspace{0.15in}
   		\includegraphics[width=0.8\textwidth]{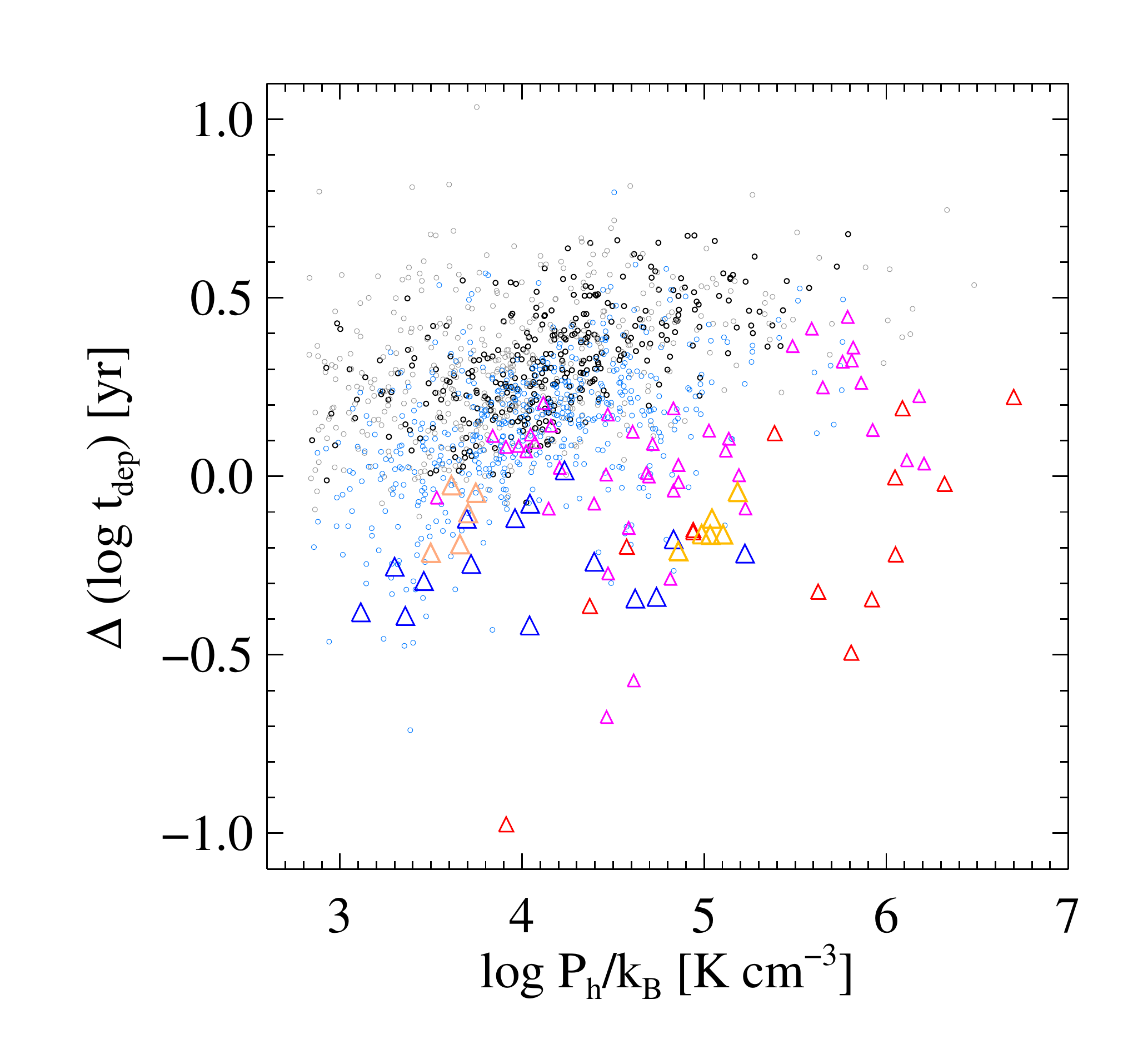}
  	\end{minipage} }
	\caption{
	Residuals from the t$_{dep}-$ sSFR relation as a function of $\Sigma_*$ (top panel),
    $\rm \Sigma_{H_{2}}$ (middle panel), and midplane gas pressure (bottom panel).
    Grey and black circles are the grids in and outside the arm regions of the spiral 
    galaxies from HERACLES. Light blue circles denote grids from galaxies without 
    spiral-arms. Blue and yellow triangles denote the grids in the bars and the ring. 
    Beige triangles denote the grids from the ring galaxy but outside the ring region.
	Cyan and green stars represent the ATLAS$^{3D}$ and COLD GASS sample respectively. } 
\end{center}
\end{figure}

Motivated by these results, we fit a plane to the two-dimensional relation between the depletion time,
specific star formation rate and the stellar surface density. Following
the method in Bernardi et al. (2003),  we find the best-fit  linear correlation of the form  
\begin {equation} t_{dep} = a\log(sSFR) +b\log(\Sigma_{*}) + c \end {equation} 
We plot all the data, including the grids from HERACLES, the ATLAS$^{3D}$ sample 
and the COLD GASS sample in Figure 13, and we show  
the t$_{dep}-$sSFR relation in the next panel for comparison. The 1:1 relation
is shown by a black solid line. The values of the coefficients, a, b, and c from the
best-fit result are $-$0.36, $-$0.14, and 5.87 respectively.
The Pearson linear correlation coefficient r between the $x$ and $y$ axes is 0.68, and
the scatter in t$_{dep}$ about the plane is 0.18 dex. 
The reduction in scatter, compared to the 
one-dimensional relation for t$_{dep}$ versus sSFR, which has a scatter of 0.19,
is only  5\%. This is because our sample is heavily dominated by the number of grids from 
the disc-dominated regions of galaxies. The main effect of the two-parameter fit is to 
bring the outlying points, which are
mainly from bulge and bar grid cells, into better agreement with the rest of the sample.  

Finally, we note  that for 1 kpc$^2$ grid cells, we can rewrite equation (3) above as
\begin {equation} t_{dep} = a\log(\Sigma_{\rm SFR})+(b-a)\log(\Sigma_{*}) + c \end {equation}
This means that the surface density of evolved stars in a grid cell contributes $\sim$1.4 times
more weight to our prediction of depletion time than the surface density of newly formed
stars! We will speculate on possible reasons for this in the next section.

\section{Summary and Discussion}     
Using the resolved maps of the nearby HERACLES galaxies, we have calculated 
the molecular gas depletion time in 1-kpc $\times$ 1-kpc grid cells
and studied the t$_{dep}-$sSFR relation for  grids lying within 
different galaxy structures such as  bulges, arms, bars, and rings.

Our main results can be summarized as follows:

(i) Bulge regions have shorter gas depletion times than the disk regions at
a given value of the  sSFR. This effect is strongest in the central regions of the bulge.

(ii) The t$_{dep}-$sSFR relation for the grid cells in spiral-arm galaxies shifts to  
longer t$_{dep}$ at fixed sSFR compared to grid cells from galaxies without spiral arms.
 Note that this shift applies to  {\em all grid cells} from spiral arm
galaxies. Within a spiral arm, the sSFR values are higher than in the inter-arm
region, so the average depletion times will be shorter.  

(iii)Grid cells located within bars have reduced depletion time at fixed sSFR.              
Grids located in barred galaxies outside of the barred region occupy  
a narrow range of low sSFR values.

(iv) We identify the stellar surface density $\Sigma_*$ as the parameter that can
best predict the shifts between the relations found for bulges, bars and spiral arms. 

\bigskip
Our results on star formation efficiencies in spiral arms are largely 
consistent with those presented in Figure 7 of Foyle et al. (2010) and 
Figure 13 of Rebolledo et al. (2013) even though our methodology for 
identifying spiral arms differs from these papers.  
Several previous papers have claimed that star formation efficiencies in
bars are low (e.g., Reynaud \& Downes 1998; Momose et al. 2010 ; 
Sorai et al. 2012). We note that in our sample, the molecular gas 
depletion time in barred regions spans a range of more than a factor of 10. 
Only in bars with high gas densities do we find grid cells with short depletion 
times and high SFEs. This is consistent with the work of Sheth et al. (2005), 
who find that barred spiral galaxies generally have higher molecular gas 
concentrations in their central regions. However, barred galaxies exist that
have no molecular gas detected in the nuclear region
and very little within the bar corotation radius.
Bars are long-lives, and we likely observe them at different evolutionary
stages, which explains why the SFE has such a large spread.

Our result that bulge grids have reduced depletion time 
compared to disc grids is consistent with results presented in Leroy et al. (2013).  
These authors studied radial trends in molecular
gas depletion time as a function of radius and observed that the average 
$t_{dep}$ was smaller in the central $\sim$1kpc regions. The reduced depletion time 
was even more pronounced after they applied a CO-to-H$_2$ conversion factor that 
depended on dust-to-gas ratio. Once again, these authors did not study trends of depletion
time at fixed specific star formation rate. 

We note that Davis et al. (2014) found   
that star formation efficiencies (SFEs) of the ETGs in ATLAS$^{3D}$ survey 
were {\em reduced} compared to  spiral galaxies.
The main reason for this apparently discrepant result
is the definition of star formation efficient adopted by these authors: SFE was computed  
as the ratio of the total (atomic+molecular) gas mass divided by the total star formation rate
of the galaxy. The analysis in this paper only considers the
star formation efficiency of the molecular gas component of the galaxy.
We note that {\em no previous study has considered variations
in SFE at fixed specific star formation rate}, which is
the main focus of this paper

Why should the star formation efficiency of the molecular gas be so sensitive to the local
surface density of evolved stars? Helfer \& Blitz (1993) studied the properties of the
dense molecular gas in the bulges of a sample of 19 normal spiral galaxies. They observed the
3mm emission from the molecules HCN and CS, which traces much denser gas than CO. They found
that the CS to CO ratio in the external bulges was consistent with that measured for the
bulge of the Milky Way and a factor of 2 or more larger than in the Galactic disc. In 
subsequent work, Helfer \& Blitz (1997) carried out interferometric observations 
of emission from the HCN molecule in NGC 6946, NGC 1068, and the Milky Way and found that 
the ratio of HCN to CO at the galactic centres was  5-10 times higher than
in the discs of these galaxies.

Helfer \& Blitz (1993) put forward two possible explanations for their findings:
1) Because the stellar potential is larger in the bulge than in the disc, the gas is subject
to much higher pressures, 2) The giant molecular clouds (GMCs) have different mass and/or 
structure in the bulge than in the disc. Launhardt, Zylka \& Mezger (2002) emphasized 
that most of the ISM in the central bulge of the Milky Way is likely
in the form of very dense, compact clouds. These conclusions arise from the lack of
a significant diffuse component in high resolution submm maps of the Galactic centre.
They speculate that the physical reason for the observed clumpiness of the ISM is the
tidal stability limit -- clouds have to be dense enough to be stabilized against tidal
forces by their own gravitation.

The results in this paper suggest that there may be a {\em continuum} of molecular cloud
properties set by the local stellar surface density. If so, this would have very interesting
consequences for our understanding of galaxy evolution. Gas-to-evolved star ratios are much
higher in the early Universe than they are today (Tacconi et al 2010) and star formation
may likewise be proceeding in a quite different manner inside GMCs.

\begin{figure*} 
\begin{center}
  \includegraphics[scale=0.41]{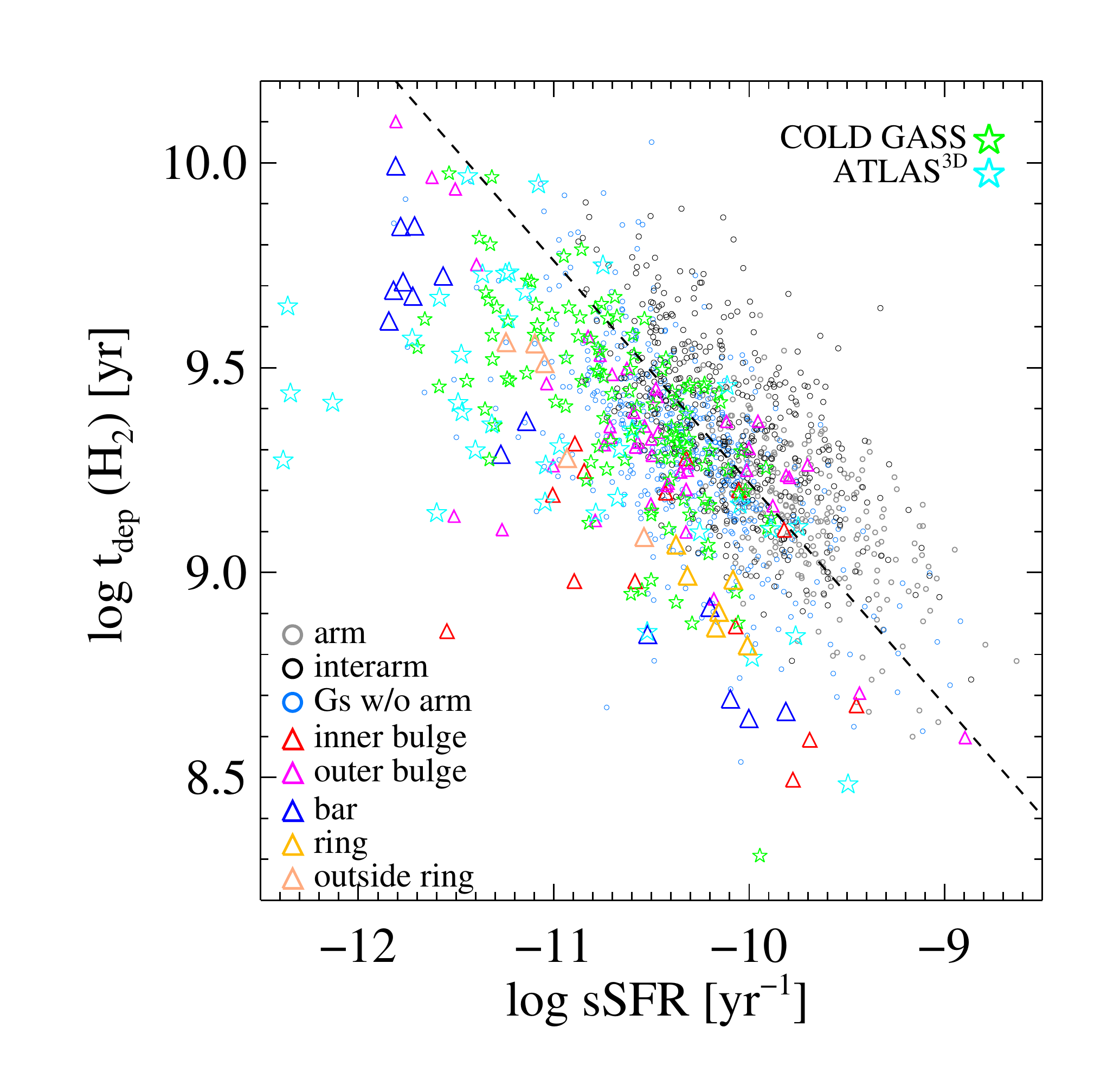}
  \includegraphics[scale=0.41]{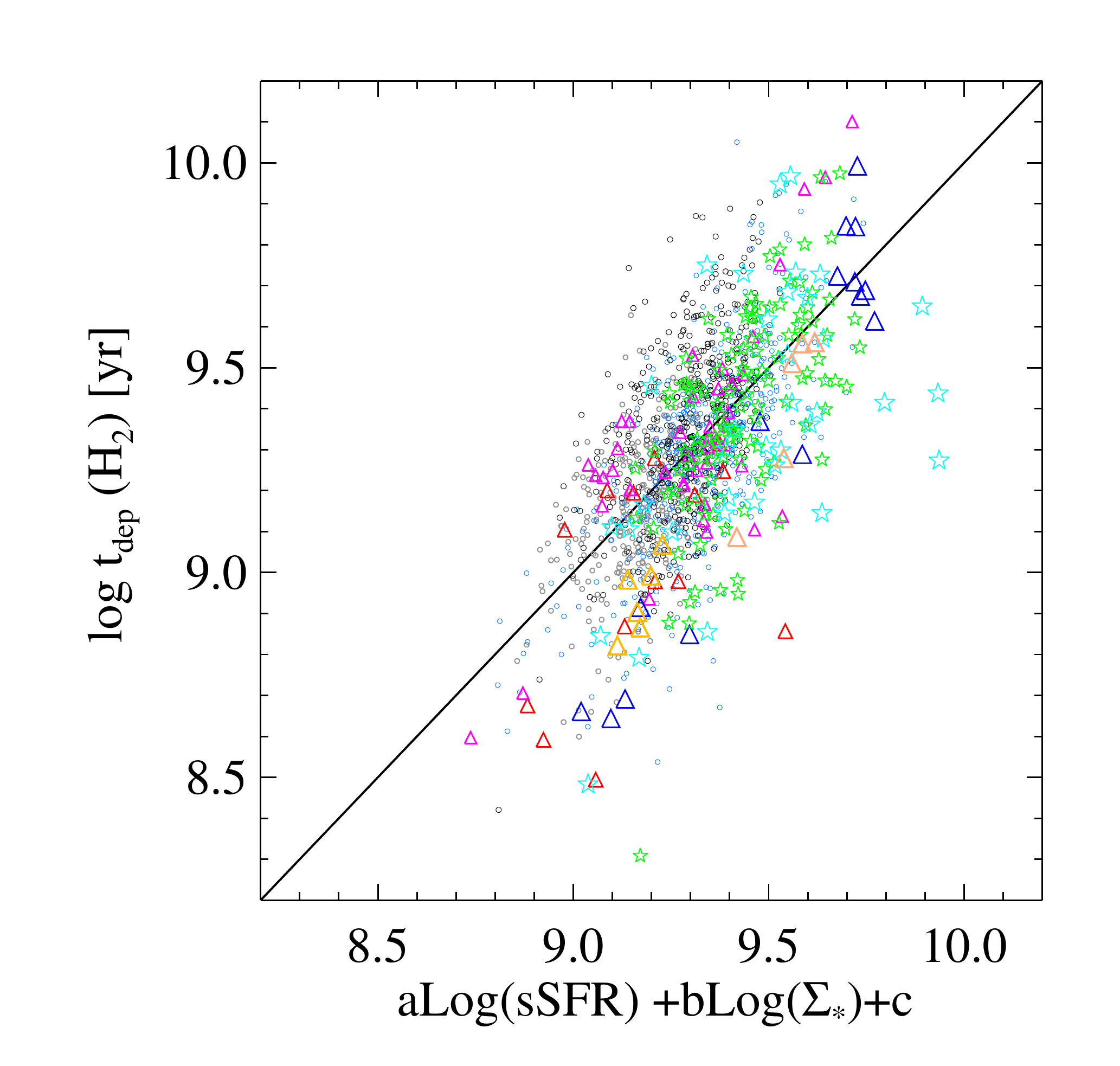}
  \caption{Left panel: t$_{dep}-$ sSFR relation for all grids of the HERACLES sample 
  ,ATLAS$^{3D}$ galaxies, and the COLD GASS galaxies. Grey and black circles are 
  the grids in and outside the arm regions of the spiral galaxies from HERACLES. 
  Light blue circle denote grids from galaxies without spiral-arms.
  Blue and yellow triangles denote the grids in  bars and rings. Beige triangles
  denote the grids from the ring galaxy but outside the ring region.
  Cyan and green stars represent the ATLAS$^{3D}$ and COLD GASS sample separately.  
  We show the OLS bisector fit to all grid data from the HERACLES sample with a black dashed line.
  Right panel: The best-fit plane linking t$_{dep}$, sSFR, and
  $\Sigma_{*}$. Symbols are as in the left plot.
  The best-fit coefficients,
  a, b, c,  are $-0.36$, $-0.14$, $5.87$ respectively.} 
\end{center}
\end{figure*}

\section*{Acknowledgments}
We thank Linda Tacconi, Reinhard Genzel, Jing Wang, Richard D'Souza, 
Sambit Roychowdhury for helpful discussions.

GALEX is a NASA Small Explorer, launched in 2003 April, developed in cooperation with 
the Centre National d'Etudes Spatiales of France and the Korean Ministry of Science 
and Technology.

This research has made use of the NASA/IPAC Infrared Science Archive, which is operated 
by the Jet Propulsion Laboratory, California Institute of Technology, under contract with 
the National Aeronautics and Space Administration.

Funding for the SDSS and SDSS-II has been provided by the Alfred P. Sloan Foundation, 
the Participating Institutions, the National Science Foundation, the US Department of 
Energy, the National Aeronautics and Space Administration, the Japanese Monbukagakusho, 
the Max Planck Society, and the Higher Education Funding Council for England. The SDSS 
is managed by the Astrophysical Research Consortium for the Participating Institutions. 
The Participating Institutions are the American Museum of Natural History, Astrophysical 
Institute Potsdam, University of Basel, Cambridge University, Case Western Reserve 
University, University of Chicago, Drexel University, Fermi National Accelerator 
Laboratory, the Institute for Advanced Study, the Japan Participation Group, Johns 
Hopkins University, the Joint Institute for Nuclear Astrophysics, the Kavli Institute 
for Particle Astrophysics and Cosmology, the Korean Scientist Group, The Chinese 
Academy of Sciences (LAMOST), the Leibniz Institute for Astrophysics, Los Alamos 
National Laboratory, the Max-Planck-Institute for Astronomy (MPIA), the 
Max-Planck-Institute for Astrophysics (MPA), New Mexico State University, 
Ohio State University, University of Pittsburgh, University of Portsmouth, 
Princeton University, the US Naval Observatory and the University of Washington.

\label{lastpage}

\end{document}